\documentclass[11pt,a4paper]{article}

\usepackage[utf8]{inputenc}
\usepackage[T1]{fontenc}
\usepackage{mathpazo}

\usepackage{amsmath,amssymb}

\usepackage{graphicx}
\usepackage{booktabs}
\usepackage{array}
\usepackage{multirow}
\usepackage{tabularray}
\usepackage{longtable}
\UseTblrLibrary{booktabs}
\UseTblrLibrary{siunitx}

\usepackage{geometry}
\usepackage{setspace}
\usepackage{caption}
\usepackage{subcaption}
\usepackage{float}
\usepackage{xcolor}
\definecolor{darkred}{rgb}{0.55,0,0} 

\usepackage[title]{appendix}
\usepackage[natbibapa]{apacite}

\usepackage[
colorlinks=true,
citecolor=teal,
linkcolor=teal,
urlcolor=teal
]{hyperref}

\usepackage{etoolbox}
\usepackage{titling}
\usepackage{codehigh}
\usepackage[normalem]{ulem}

\AtBeginEnvironment{abstract}{\singlespacing}

\NewTableCommand{\tinytableDefineColor}[3]{\definecolor{#1}{#2}{#3}}

\newcounter{bibnum}
\let\originalbibitem\bibitem
\renewcommand{\bibitem}[2][]{%
  \stepcounter{bibnum}%
  \originalbibitem[#1]{#2}%
  \arabic{bibnum}.\enspace\ignorespaces
}

\newtheorem{fnd}{Finding}

\title{\Large Playing Against the Machine: Cooperation, Communication, and Strategy Heterogeneity in Repeated Prisoner’s Dilemma}

\author{
Chowdhury Mohammad Sakib Anwar\thanks{
Department of Finance and Computer Science, University of Winchester, SO22 4NR, Winchester, U.K., Email: \href{mailto:sakib.anwar@winchester.ac.uk}{sakib.anwar@winchester.ac.uk}, 
ORCID: \href{https://orcid.org/0000-0002-0223-7963}{0000-0002-0223-7963}
}
\and
Konstantinos Georgalos\thanks{
Department of Economics,
Lancaster University Management School,
LA1 4YX, Lancaster, U.K., Email: \href{mailto:k.georgalos@lancaster.ac.uk}{k.georgalos@lancaster.ac.uk}, 
ORCID: \href{https://orcid.org/0000-0002-2655-4226}{0000-0002-2655-4226}\newline
We are grateful to the University of Winchester for providing the funds for this study. We would like to thank Michalis Drouvelis and Nicolaus Teng for their help running some of the experimental sessions and Fabian Dvorak and Sebastian Fehrler for sharing their data with us. The study has received ethics approval from the FASS/LUMS Ethics Committee at Lancaster University, FASSLUMS-2024-4251-LeXeL-3.  The study has been pre-registered at the OSF registry (\url{https://osf.io/e4rz8}). The replication material for the study is available at \url{https://osf.io/e4rz8} 
\newline \textbf{Use of Artificial Intelligence (AI) Tools} : During the preparation of this work, the authors used Clause Opus 5.5 to check grammar, refine arguments, verify code accuracy, and assist in testing the text classifier. The authors thoroughly reviewed and validated all outputs and take full responsibility for the final content.
\newline\textbf{Conflict of Interest}: None }
}

\date{\today}

\begin{document}

\maketitle

\begin{abstract}
 We study whether communication sustains cooperation when the partner is an artificial intelligence. In a laboratory experiment ($n = 234$), human participants play indefinitely repeated Prisoner's Dilemma games against an AI chatbot (GPT-5.2) under one of three conditions: no communication, free-form chat before the first round of each supergame, or free-form chat before every round. A fourth treatment replaces the AI with another participant under the every-round protocol. Communication is the main determinant of cooperation with the AI, and a single exchange is enough: pre-play chat raises cooperation from 49\% to 84\%, while chatting before every round adds less than one percentage point, in contrast to the timing effect found between humans. Once communication is available, cooperation with the AI is no lower than with a human partner, so we find no evidence of the ``machine penalty'' documented in the psychology literature. The partner's identity nonetheless shifts strategies: under identical communication, participants tend to reciprocate against a human but to cooperate unconditionally against the AI. Participants also keep their promises to cooperate while carrying out their threats to punish only about half the time. The AI opens cooperatively and rarely defects first, so these differences originate on the human side.
 
\medskip
\noindent\textbf{Keywords:} Prisoner's Dilemma, human--AI interaction, communication, cooperation, large language models, laboratory experiment

\medskip
\noindent\textbf{JEL Codes:} C72, C91, C92, D83
\end{abstract}


\newpage
\section{Introduction}
\label{sec:introduction}

AI agents are increasingly working alongside humans in roles that rely on sustained cooperation. LLM-based chatbots already serve as negotiation counterparts, financial advisors, customer-service representatives, and collaborative assistants in settings where trust, reciprocity, and repeated exchange are central to the quality of outcomes.\footnote{\citet{yang2025} provide field-experimental evidence that human--AI collaborative advice improves investment decisions, with emotional trust as the key mediator. For a broader discussion of how generative AI may reshape economic interactions and socioeconomic inequalities, see \citet{capraro2024}.} The psychology literature shows that people apply social norms such as reciprocity even to simple computers \citep{nass2000}, yet they often respond less favourably to machines than to other people \citep{qin2025, bonnefon2024}. Failed cooperation is costly, as advisory relationships collapse, bargaining stalls, and the promise of human--AI collaboration goes unrealised.

Two strands of recent work treat LLMs as objects of economic analysis. One applies revealed-preference tools to LLM behaviour, asking whether their choices satisfy standard rationality axioms \citep{caplin2025jet, autor2025}. The other uses LLMs as simulated subjects: \citet{horton2026} propose treating them as ``Homo silicus" whose responses to experimental manipulations can generate economic hypotheses at low cost. Our approach differs from both, in that we study humans interacting with an LLM rather than the LLM in isolation. We also differ from much of the psychology literature on how people respond to AI, which has relied largely on stated attitudes such as trust, liking, and perceived fairness \citep{qin2025}, whereas we observe incentivised behaviour.

Despite the rapid deployment of such agents, the most powerful mechanism for sustaining cooperation between humans, namely communication, remains almost entirely untested in human--AI strategic interaction. In human--human interactions, the evidence is unambiguous: communication is one of the strongest facilitators of cooperation in social dilemmas \citep{balliet2010, jin2025}. Communication operates through coordination on cooperative equilibria, psychological commitment via promises, and renegotiation following breakdowns \citep{crawford1982, farrell1987, charness2006, cooper2023}.\footnote{For a survey of the cheap talk literature, see \citet{farrell1996}.} Early work in social psychology similarly found that allowing a group to discuss the dilemma sharply reduces defection \citep{dawes1977}, largely because discussion elicits commitments to cooperate that people then honour \citep{kerr1994}. Modern LLMs can engage in fluent natural language conversation, yet whether that fluency is sufficient to activate these channels is an open question that is both theoretically and practically important.


This paper asks whether communication retains its cooperation-sustaining role when the opponent is an AI agent, or whether the mechanism breaks down. We investigate this question in indefinitely repeated Prisoner's Dilemma games, the setting where these mechanisms are most relevant, by varying communication against an AI partner (none, a single exchange before each supergame, or an exchange before every round) and, holding the repeated protocol fixed, varying the partner between an AI and another human, following the design of \citet{dvorak2024}. Beyond aggregate cooperation rates, we examine two dimensions through which communication effects may differ: the strategies humans adopt, which we recover using structural estimation, and the content of the conversations themselves, including whether humans alter their use of coordination messages, promises, and threats when addressing an AI. These questions sit at the intersection of three strands of the literature: the role of communication in repeated games, human responses to AI partners, and the strategic behaviour of LLMs.

The experimental evidence on communication in repeated games, and in particular \citet{dvorak2024}, whose design we adopt, is the most closely related work to ours. They vary communication timing in indefinitely repeated Prisoner's Dilemma (PD) games under a factorial design. They find that pre-play communication increases cooperation by 53 percentage points under perfect monitoring, and that every-round communication yields additional gains, primarily by allowing participants to exchange messages and revisit incomplete agreements, a timing effect that is absent in the broader meta-analytic evidence where pre-play and continuous communication produce statistically indistinguishable effects \citep{balliet2010}. In their design actions are observed through signals that are correct with probability 0.9 even under perfect monitoring, whereas ours are noiseless, so we compare treatment differences rather than levels. The key channel is coordination messages: explicit proposals for mutual cooperation that reduce strategic uncertainty. Critically, strategy estimation reveals that under every-round communication with perfect monitoring, 100\% of subjects adopt lenient, forgiving strategies (94--100\% across monitoring conditions), suggesting that ongoing communication fundamentally changes not just the level but the nature of cooperative behaviour. More broadly, the experimental literature has shown that cooperation increases with experience \citep{dal2011, dal2019}, is higher when cooperative equilibria are risk-dominant \citep{blonski2011}, and converges on conditionally cooperative strategies, predominantly Grim Trigger and Tit-for-Tat \citep{axelrod1984}, as documented through both structural estimation \citep{dal2011} and direct observation of team discussions \citep{cooper2023}. These regularities characterise human--human repeated-game behaviour and provide the reference point against which we assess whether human--AI interaction produces different strategic patterns.


Whether these patterns carry over to human--AI interaction is, however, uncertain. \citet{chugunova2022}, in a review of 138 papers, identify reduced emotional responses, less mentalising, and lower cooperation against machines, with the distribution of agency emerging as a key moderating factor. On the question of why these patterns arise, \citet{burton2020} organise algorithm aversion into five themes and highlight the role of the expectations and beliefs that users form about algorithmic reasoning.\footnote{For an earlier review focused on strategic interactions with computer players, see \citet{march2021}. Algorithm aversion extends beyond strategic settings; for example, \citet{dargnies2024} document aversion to hiring algorithms even when they outperform human evaluators.} Consistent with these broader patterns, several papers find that cooperation falls when participants know they face an AI \citep{dvorak2025, engel2025, vonschenk2025}, that social preferences toward machines differ systematically from those toward humans even when payoff structures are identical \citep{vonschenk2025}, and that humans may shift from perspective-taking to rule-based reasoning when paired with an AI partner \citep{bayer2024}.\footnote{\citet{almog2025} provide complementary evidence from professional tennis, showing that AI oversight systematically alters human decision patterns.} At the same time, there is evidence that communication capacity can bridge this gap: \citet{crandall2018} demonstrate that even pre-programmed speech acts approximately double mutual cooperation rates, bringing human--machine cooperation to levels comparable with human--human pairs, suggesting that communication capacity, not just learning sophistication, may be the critical ingredient for machine cooperation.\footnote{Related work points to a more nuanced picture, with AI-generated promises being used more frequently but believed less individually, with overall trust and efficiency remaining similar to human communication \citep{bogliacino2023}, transparency about algorithms involving efficiency tradeoffs \citep{ishowo2019}, and human--AI strategic dynamics evolving substantially over repeated play \citep{werner2026}.} The emerging consensus is that human--machine systems must be studied as complex social systems whose collective outcomes cannot be deduced from either side alone \citep{tsvetkova2024, rahwan2019}.\footnote{Evidence of behavioural spillovers from human--bot to subsequent human--human interactions further underscores the need for a systems-level approach \citep{spillovers2025}.}

The psychology literature documents this reluctance in detail. In a meta-analysis of 163 studies, \citet{qin2025} show that people favour an AI over a human only when they perceive it as more capable and see no need for personalised treatment, and are averse otherwise, with appreciation weaker when behaviour rather than attitude is measured. In incentivised games, cooperation with machines suffers what \citet{bonnefon2024} call a ``machine penalty'', clearest in one-shot games without communication, where people expect an AI partner to be as cooperative as a human, yet cooperate with it less and feel less guilt about exploiting it \citep{karpus2021}. Nor is it obvious that conversation helps, since a response that makes people feel heard loses part of its effect once it is known to come from an AI \citep{yin2024}. Whether conversation can overcome the machine penalty in repeated, incentivised interaction is what our design tests.


A closely related question concerns the strategic behaviour of modern LLMs themselves. Recent papers find that LLMs cooperate at rates substantially higher than those of humans across a range of social dilemma games \citep{engel2025, fontana2025nicer, mei2024, xie2024}, with behaviour well described by conditional welfare maximisation \citep{bauer2025}.\footnote{Early evidence on LLM behaviour in games appears in \citet{guo2023} and \citet{brookins2024playing}; for evidence specifically from repeated PD with an LLM partner, see \citet{orland2025}. On the broader question of whether LLMs can simulate human experimental participants, see \citet{aher2023} and \citet{horton2026}.} Closest on the algorithmic side is \citet{kasberger2023}, who compare Q-learning algorithms with human play in the indefinitely repeated PD. Varying the continuation probability and the reward from cooperation, they find that the same economic factors govern cooperation for both, although algorithms cooperate less than humans when cooperation is risky, and that the strategies differ. Using the Strategy Frequency Estimation Method they show that, under partner matching, algorithms rarely resort to \emph{grim trigger} and instead sustain cooperation through \emph{win-stay-lose-shift} and a cycling \emph{win-shift-lose-shift} strategy that humans hardly ever use. Most relevant for our purposes, an earlier version of their paper reports an experiment in which two GPT-3.5 agents (\texttt{gpt-3.5-turbo-0301}) play the same game against each other \citep{kasberger2023wp}. The LLMs cooperate at high rates and adopt mostly cooperative strategies such as unconditional cooperation and \emph{tit-for-tat}, yet their cooperation rates are largely unresponsive to the continuation probability and the reward parameter. These cooperative tendencies come, however, with important fragilities, as \citet{akata2025} find that GPT-4 tends to be unforgiving after a single defection and does not respond to changes in continuation probability, the parameter that most strongly governs human cooperation in repeated games. Taken together, these traits mean that an LLM partner behaves very differently from a human partner, so communication may work through different channels or fail entirely. Crucially, however, none of the papers reviewed above examines what happens when humans can engage in natural language conversation with an AI partner in a repeated strategic setting.\footnote{Concurrent work by \citet{cooperative2025} shows that communication increases cooperation with LLMs in one-shot settings, and \citet{experimental2025} document broadly cooperative human–LLM interactions in finitely repeated games, but without pre-play communication or indefinite repetition, the mechanisms central to our design.}


We address this gap with what is, to our knowledge, the first controlled experiment in which humans engage in natural language conversation with an AI partner across indefinitely repeated strategic interactions, extending the communication--cooperation literature \citep{balliet2010, dvorak2024} beyond its exclusively human--human domain.
Our design separates three effects. A no-communication baseline against the AI identifies the effect of communication itself. Comparing pre-play with repeated communication tests whether the timing effects documented by \citet{dvorak2024} in human--human interactions extend to human--AI settings, thereby distinguishing initial agreement formation from ongoing renegotiation. A human--human treatment under the repeated protocol, run with the same software and game parameters, identifies the effect of the partner's identity without relying on a cross-study comparison. We further analyse the content of the conversations, building on \citet{penczynski2019} and \citet{cooper2023} in linking communication content to strategic behaviour. We code every message with a codebook whose categories are the strategies of the repeated game itself, implemented as a transparent rule-based classifier and validated against independent human coding \citep{BaranskiCooperLee2026, CelebiPenczynski2026}, and we link each communicated rule to the decision that follows it. Finally, we employ the Strategy Frequency Estimation Method \citep{dal2011} to characterise the distribution of strategies humans adopt against an AI partner, an important question given evidence that the benefits of AI collaboration depend on individual calibration \citep{caplin2024abc} and that the strategy distributions of LLMs are notably more concentrated than those of humans \citep{mei2024, xie2024}.


We conduct a laboratory experiment in which 234 human participants play indefinitely repeated PD games against an AI chatbot powered by GPT-5.2 or against another human participant. Our between-subjects design features four treatments: a \emph{HAI-Rep} treatment where participants converse with the AI before every round, a \emph{pre-play chat} treatment where conversation occurs only before the first round of each supergame, a no chat treatment where participants play the AI without any communication, and a human–human treatment where participants converse with another participant before every round. The design closely follows \citet{dvorak2024}, adopting their payoff parameters ($T = 37$, $R = 30$, $P = 17$, $S = 0$), supergame structure (7 supergames with $\delta = 0.80$ continuation probability), and length-generation methodology, which allows us to ask whether the communication effects they document between humans carry over to a human--AI setting. The effect of the partner's identity, by contrast, is identified within our own experiment, by the human--human treatment. Our signals are perfectly accurate, whereas theirs are correct with probability 0.9, so we compare the two studies in terms of treatment effects rather than cooperation levels.


We find four main results. First, communication is the main determinant of cooperation against an AI partner, and a single exchange is enough. Measured by the share of rounds in which the participant cooperates, cooperation is 49\% without communication; a pre-play exchange raises it to 84\%, an effect that is present from the first round of each supergame and does not fade with experience. Allowing further exchanges before every round adds less than one percentage point (85\%), whereas between humans \citet{dvorak2024} find that repeated communication raises cooperation further; our null timing effect is closer to the meta-analytic evidence of \citet{balliet2010}. A plausible explanation lies in the behaviour of the AI. It opens with cooperation in at least nine out of ten supergames in every treatment, so the renegotiation function that gives within-supergame communication its value between humans has no work to do.

Second, once communication is available, cooperation against an AI is no lower than against another human. Holding the repeated-communication protocol fixed, participants cooperate about 8 percentage points more with the AI than with a human. The gap is present from the first round, is at the margin of conventional significance overall,  but is no longer statistically distinguishable from zero in the last three supergames. This contrasts with the evidence that cooperation falls when participants know they face an algorithm \citep{dvorak2025,engel2025,vonschenk2025}, the ``machine penalty'' of \citet{bonnefon2024}, and is consistent with \citet{crandall2018}, who find that communication capacity is what brings human–machine cooperation to human levels.

Third, using strategy frequency estimation, we find that the identity of the partner shifts the strategies humans adopt even when the communication protocol is held fixed. Unconditional defection and Grim Trigger dominate without communication. Once communication is introduced, unconditional defection disappears, with the mass concentrating on Grim Trigger under pre-play communication and on unconditional cooperation under repeated communication. Under the identical repeated protocol, the modal strategy against a human is Tit-for-Tat (40\%), which is absent against the AI, while the modal strategy against the AI is unconditional cooperation (40\%, against 15\% with a human partner). When they can talk every round, humans therefore tend to reciprocate against a human but to cooperate unconditionally against an AI, the opposite of the rule-based caution one might expect from the algorithm aversion literature \citep{bayer2024}. Because cooperation is so high, few histories leave the cooperative path, so these estimates describe play that is consistent with a strategy rather than cleanly separating unconditional from conditional cooperation. The AI itself is best described by Tit-for-Tat in every treatment, with a sizeable Grim Trigger share only under pre-play communication; it almost never initiates defection without chat and rarely with it, and once defected upon it punishes for longer than humans do, consistent with the unforgiving behaviour documented by \citet{akata2025}. When cooperation breaks down, it is almost always the human who defects first.

Fourth, conversation with an AI is deliberative and is implemented, whereas conversation with a human is a coordination signal.
Under the same repeated protocol, human messages to the AI contain more payoff reasoning and more talk about credibility than human messages to another human, which are short and almost entirely rule-free. Explicit repeated-game rules are a feature of pre-play chat, where one in five human statements sets one out; when chat is available every round they are rare against an AI and a human alike (about 3\%). Participants largely do what they say, but not symmetrically. Cooperative commitments are almost always honoured, whereas punishment threats are carried out in only about half of the cases in which the partner actually defects. Social and affective content is not lower when the partner is an AI.

These results show that humans do not need ongoing conversation to cooperate with an AI. A plausible reason why one exchange is enough is that the AI is predictable. It opens cooperatively, follows the rule it states, and rarely defects first, so there is little for further conversation to repair. Communication with an AI therefore works mainly through one channel, coordination on the cooperative equilibrium.


The remainder of this paper is organised as follows. Section~\ref{sec:theory} develops the theoretical framework. Section~\ref{sec:design} describes the experimental design. Section~\ref{sec:results} presents the results. We then conclude.
\section{Theory}
\label{sec:theory}

Our theoretical framework follows \citet{dvorak2024}, who study the role of communication in the indefinitely repeated prisoner's dilemma under different monitoring structures. We restrict attention to perfect monitoring and adapt their framework to the human-AI setting.
\subsection{The Game}
A human player and an AI player interact over indefinitely many rounds of a supergame. After each round, the interaction continues with probability $\delta = 0.8$. In every round, each player simultaneously chooses $a_i \in \{C, D\}$. The stage-game payoffs in ECU are:
\[
\begin{array}{c|cc}
 & C & D \\ \hline
C & 30, \; 30 & 0, \; 37 \\
D & 37, \; 0 & 17, \; 17
\end{array}
\]
so that the reward from mutual cooperation is $R = 30$, the temptation payoff is $T = 37$, the punishment payoff is $P = 17$, and the sucker payoff is $S = 0$. Under perfect monitoring, both players observe the action profile $(a_i, a_{-i})$ at the end of each round. The history of the game is therefore common knowledge.
\subsection{Cooperative Equilibria}
The grim-trigger strategy supports cooperation in a subgame-perfect equilibrium whenever $\delta \geq \delta^{SPE} = (T-R)/(T-P) = 7/20 = 0.35$, which is satisfied by $\delta = 0.8$. Since players observe true actions, full cooperation can be sustained on the equilibrium path. Beyond pure strategies, cooperative \emph{memory-one belief-free} (M1BF) equilibria, behaviour strategies that condition only on the previous round's action profile, also exist. An M1BF strategy is a vector $\sigma = (\sigma_\emptyset, \sigma_{CC}, \sigma_{CD}, \sigma_{DC}, \sigma_{DD})$ specifying the cooperation probability after each possible history. These equilibria exist whenever $\delta \geq \delta^{BF} = (P-S)/(R-S) = 17/30 \approx 0.57$, again satisfied in our setting. At this threshold, the unique equilibrium strategy is $\sigma = (\sigma_\emptyset, 1, 10/17, 1, 0)$: certain cooperation after $CC$, certain defection after $DD$, leniency after the opponent's defection ($\sigma_{CD} = 10/17$), and full forgiveness after one's own defection ($\sigma_{DC} = 1$). A useful subclass are \emph{semi-grim} strategies, which impose symmetry across the two asymmetric histories: $\sigma_{CD} = \sigma_{DC}$. \citet{breitmoser2015cooperation} shows that semi-grim strategies describe experimental behaviour well under perfect monitoring. These exist whenever $\delta \geq \delta^{SG} = \left[(T-R)+(P-S)\right]/(T-S) = 24/37 \approx0.65$, satisfied by $\delta = 0.8$. At the threshold, the unique strategy is $\sigma = (\sigma_\emptyset, 1, 17/24, 17/24, 0)$. Renegotiation-proof equilibria, relevant when players can communicate, also exist under our parameters \citep[Appendix A.3.1]{dvorak2024}.

In summary, the parameterisation supports a rich set of cooperative equilibria. It also supports mutual defection, which is an equilibrium of any repeated prisoner's dilemma. The existence of cooperative equilibria therefore does not guarantee cooperation: which equilibrium is selected depends on what each player expects the other to do, and it is this expectation that communication is able to shape. Whether it does so when the partner is an AI is the question our treatments address.

\subsection{Communication, Research Questions, and Predictions}
\citet{dvorak2024} identify two roles for communication: (i) reducing strategic uncertainty by coordinating on cooperation, and (ii) reducing uncertainty about the history of play. Under perfect monitoring, only the first role is relevant, since both players observe true actions and the history is common knowledge. Their human-human results confirm this: pre-play communication is sufficient for high and stable cooperation under perfect monitoring, with repeated communication offering only a modest additional benefit. Note that in \citet{dvorak2024} actions are translated into signals that are correct with probability 0.9 and
payoffs follow the signal even under perfect monitoring, whereas in our experiment actions
are observed and paid without noise.
 
Our experiment asks whether these findings carry over when one player is replaced by an AI and signals are perfectly accurate. The equilibrium structure is unchanged: the same parameters support the same set of cooperative equilibria regardless of whether the opponent is human or AI. However, whether humans behave differently when facing an AI is a behavioural question that the standard theory does not address. We organise our investigation around four research questions.
 
\medskip

 \noindent \textbf{Question 1:} \emph{Does communication retain its cooperation-sustaining role when the opponent is an AI, or does the mechanism break down?}
 
\noindent Under perfect monitoring, communication promotes cooperation by reducing strategic uncertainty through coordination on mutual cooperation. This mechanism should operate regardless of the opponent's type, provided that (a) the AI's stated intentions are informative about its subsequent behaviour, and (b) the human treats the AI's messages as credible. If these conditions hold, we expect communication to lead to high cooperation rates, as in the human-human benchmark.
 
\medskip
\noindent \textbf{Question 2:} \emph{Does facing an AI opponent shift the distribution of strategies humans adopt in the indefinitely repeated PD?}
 
\noindent The set of equilibria is identical to the human-human case, so the same strategies (grim-trigger, lenient and forgiving M1BF strategies, semi-grim strategies) are all viable. The theory gives no directional prediction: any shift in strategy distribution against an AI opponent would reflect behavioural rather than strategic forces, such as different beliefs about the AI's likely play or different attitudes towards cooperating with a machine.
 
\medskip
 
\noindent \textbf{Question 3:} \emph{Does facing an AI opponent change how humans communicate in terms of tone, content, and use of promises and threats?}
 
\noindent The theory predicts that communication should be used to coordinate on cooperation, but is silent on tone, affect, or the specific language used. This question is largely exploratory. However, since the mechanism through which communication sustains cooperation relies on credible agreements, it is relevant to understand whether the content and character of communication differ when the partner is an AI. The human-human treatment with repeated communication, which uses the same chat protocol, provides the benchmark for this comparison.
 
\medskip

\noindent \textbf{Question 4:} \emph{Is the effect of pre-play versus repeated communication on cooperation and strategy choice the same against an AI opponent as against a human?}
 
\noindent Under perfect monitoring, \citet{dvorak2024} find that pre-play communication is sufficient for stable cooperation and that repeated communication offers limited additional benefit. The theoretical rationale is that there is no uncertainty about the history of play to resolve through ongoing communication. This prediction carries over directly to the human-AI case. However, repeated communication may serve additional functions in the human-AI context, such as allowing the human to assess the AI's responsiveness or to build trust incrementally, which could lead to a larger gap between the two treatments than observed in the human-human setting.
 
\section{Experimental Design}
\label{sec:design}

\subsection{Treatments}
\label{sec:design_treatments}
In our experiment, each subject plays seven indefinitely  repeated Prisoner's Dilemma supergames against either an AI chatbot powered 
by the large language model GPT-5.2\footnote{We use the pinned snapshot of the reasoning model \texttt{gpt-5.2-2025-12-11} with reasoning effort set to medium, to ensure reproducibility across sessions.} or a  human opponent. Our design varies two dimensions, one at a time. Against an AI opponent we vary communication across three levels: none, a single exchange before the supergame, and an exchange before every round, holding the partner fixed. Holding communication fixed at the repeated protocol, we vary the partner between an AI and another human. This yields four treatments in total: HAI-Pre (AI opponent with communication only before the first round of each supergame), HAI-Rep (AI opponent with communication before every round of each supergame), HAI-NoChat (AI opponent without communication), and HH-Rep (human opponent with communication before every round of each supergame)\footnote{We do not run a human–human treatment without communication as the result that communication raises cooperation substantially in human–human repeated dilemmas is well established \citep{dvorak2024, andres2023, balliet2010}, and our question is whether it operates similarly when the partner is an AI.}
\paragraph{HAI-Rep.} Participants engage in free-form natural language chat with the AI chatbot before making their decision in \emph{every round} of each supergame. The chat window is available for 120 seconds, during which participants can exchange multiple messages with the AI. After the chat phase, participants submit their decision ($A$ or $B$). The AI's decision is made simultaneously, informed by the full conversation history within the current supergame and the accumulated history of play.

\paragraph{HAI-Pre.} Participants chat with the AI chatbot only before the \emph{first round} of each supergame. The initial chat window is available for 180 seconds, longer than in the repeated treatment to compensate for the single communication opportunity. In all subsequent rounds, participants proceed directly to the decision stage without chat, facing a 60-second decision timer. The AI's decisions in later rounds are informed by the initial conversation and the accumulated history of play.

\paragraph{HAI-NoChat.}
As a benchmark to compare behaviour, participants in this treatment face the exact same structure of the game as above (identical number of supergames, continuation probability, stage-game payoffs) but without the option to exchange messages with the AI opponent. 

\paragraph{HH-Rep.}
As a second benchmark, participants in this treatment face the exact same structure of the game as in the HAI-Rep treatment with the only difference that the opponent now is another subject, and the chat window was open for 40 seconds\footnote{ We allow longer chat windows in HAI to accommodate delays from server processing and API response times.}.

The design therefore identifies two things separately: the effect of communication, from the three human-AI treatments that differ only in the communication protocol, and the effect of the partner's identity, from HAI-Rep and HH-Rep, which differ only in whether the opponent is an AI or another participant. As there is no human--human treatment without communication,  we cannot identify the interaction of the two dimensions.

The experiment was implemented using oTree \citep{chen2016otree} and conducted at the Lancaster Experimental Economics Laboratory (LExEL) and the Birmingham Experimental Economics Laboratory (BEEL) between March and June 2026. Subjects were recruited using ORSEE  \citep{greiner2015}. \footnote{The HAI treatments were run at LExEL; the HH-Rep and HAI-NoChat treatments were run at BEEL, as the LExEL subject pool had been exhausted and recruitment was not feasible there during the relevant period. The two laboratories draw on
comparable undergraduate subject pools at  large UK universities and follow the same recruitment procedures, and the protocol, instructions and software were identical across sites, so we do not expect the change of location to affect behaviour.}

Sessions proceeded as follows. Upon arrival, participants were seated at individual computer terminals and read on-screen instructions describing the game, the payoff matrix, the supergame structure, and the payment rule. The description of the partner differed by treatment. In the three HAI treatments, the instructions explicitly stated that participants would be interacting with an AI chatbot powered by a large language model. In HH-Rep, they stated that participants would be interacting with another participant in the session. 

After reading the instructions, participants completed a comprehension quiz verifying understanding of key elements: the number of supergames, the continuation probability, the payoff matrix, and the payment rule. Participants had to answer all quiz questions correctly before proceeding. After the final supergame, participants completed a brief survey collecting demographic information including age, gender, degree programme, year of study, ethnicity, and religion. Sessions lasted less than 40 minutes on average.

\subsection{Game Structure and Payoffs}
\label{sec:design_structure}

Each session consists of 7 supergames.  In the HAI treatments, the AI opponent was reset at the start of each supergame, so that  participants faced a fresh AI with no memory of previous supergames.   
This mirrors the stranger re-matching protocol in  \citet{dvorak2024}, where participants were paired with a new human partner for each supergame\footnote{In order to maintain the stranger re-matching protocol the system prompt provided to GPT-5.2 did not include the number of supergames or the current supergame number, ensuring each AI instance had no knowledge of the broader session structure.}. 

The length of each supergame is determined by a stochastic process with a continuation probability of $\delta = 0.80$ per round, yielding a geometric distribution with an expected supergame length of 5 rounds. Following \citet{dvorak2024}, we pre-generate the supergame length sequence using their methodology: sequences are drawn conditional on the median length equalling 5 rounds, using a fixed random seed for reproducibility. All sessions use the same realised supergame lengths, eliminating between-session variation in game length as a confound. Participants are informed that each supergame has an 80\% probability of continuing after any given round and that there are 7 supergames in total. The exact length of each supergame is not revealed in advance.

In each round, the two players simultaneously choose action $A$ (cooperate) or $B$ (defect). Payoffs are as described in Section~\ref{sec:theory} and are denominated in experimental points. Points are converted to British pounds at a rate of \pounds 0.08 per point. At the end of the experiment, one supergame is randomly selected for payment. The participant's earnings from that supergame, plus a \pounds 5.00 participation fee, constitute their total payment.\footnote{This payment rule is standard in repeated-game experiments \citep{azrieli2018} and ensures that participants treat each supergame as payoff-relevant while keeping expected earnings manageable.} The average earnings from the experiment was \pounds14.89 (this includes the \pounds 5 show-up fee).

\section{Results}
\label{sec:results}
We first present some summary statistics on choices and cooperation rates. We then explore strategy heterogeneity with the aid of structural modelling and conclude this section with some regression analysis. 

\subsection{Descriptive Statistics}
\begin{figure}[H]
\centering
\begin{subfigure}{0.48\textwidth}
    \centering
    \includegraphics[width=\linewidth]{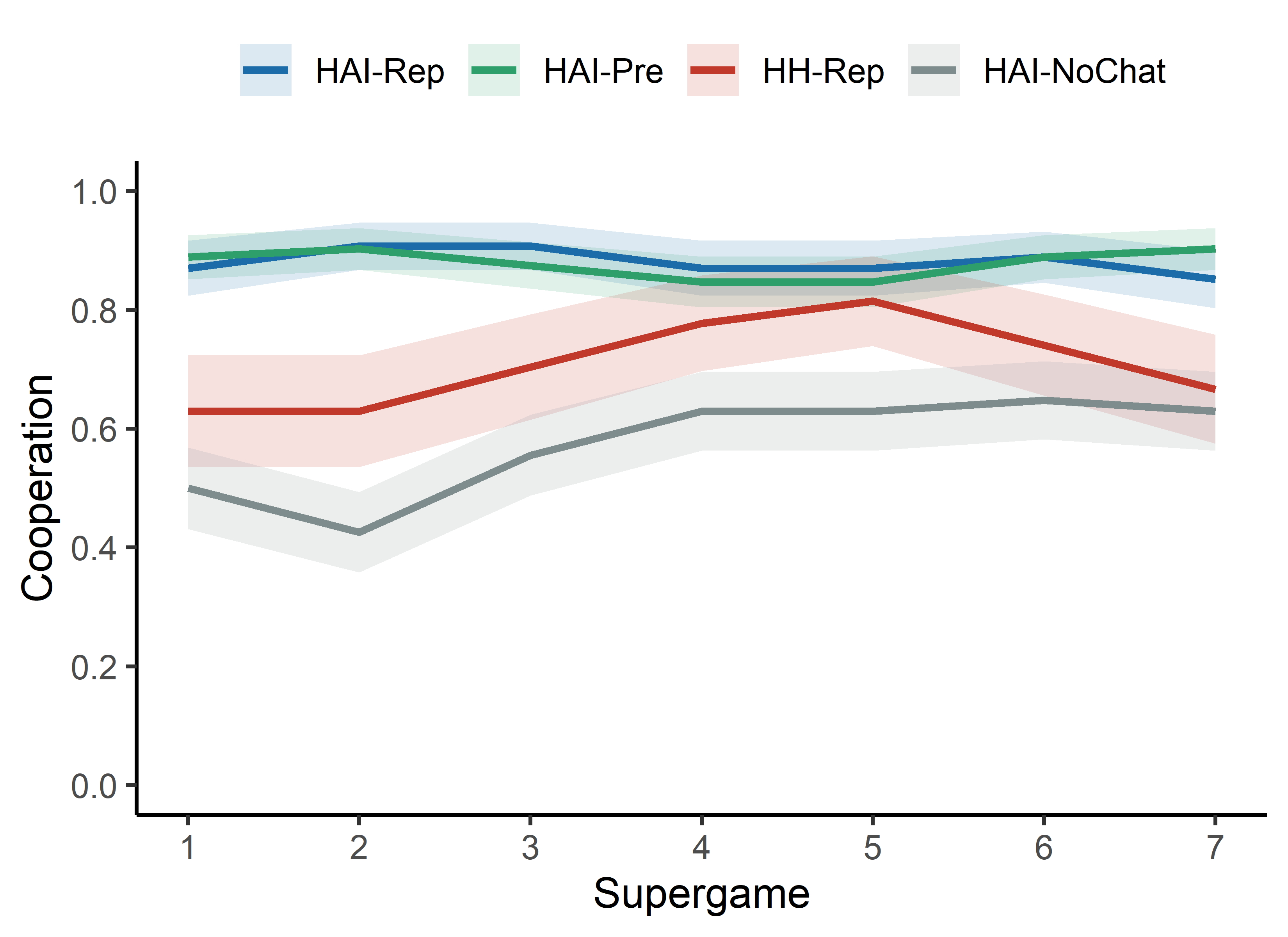}
    \caption{Round 1}

\end{subfigure}
\hfill
\begin{subfigure}{0.48\textwidth}
    \centering
    \includegraphics[width=\linewidth]{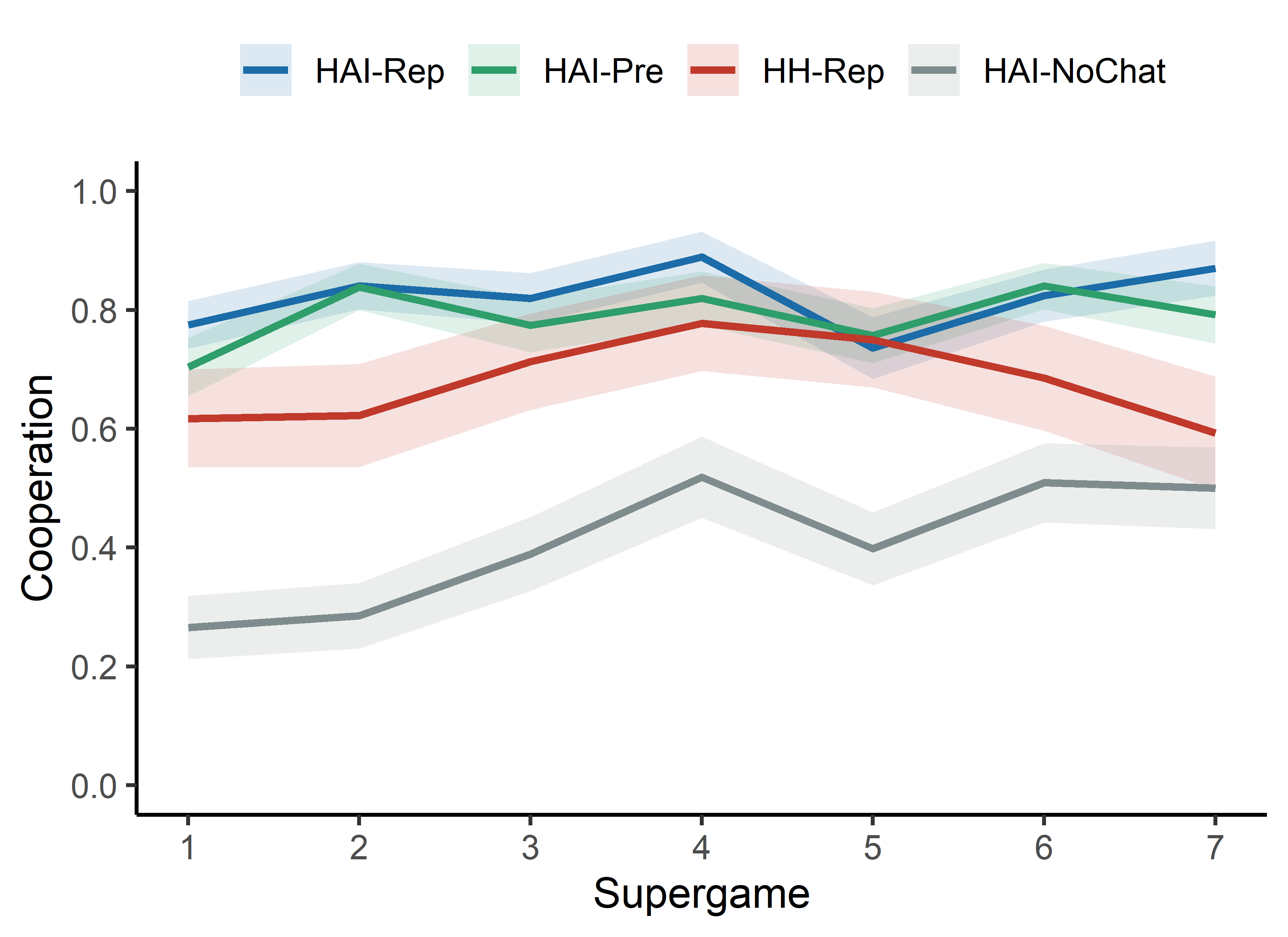}
    \caption{Rounds > 1}
    \label{fig:rep_all}
\end{subfigure}

\caption{Mean joint cooperation by supergame and treatment. The left panel reports round 1 of each supergame; the right panel reports all subsequent rounds. Shaded areas denote ±1 standard error of the mean.}
\label{fig:coop}
\end{figure}

Figure~\ref{fig:coop} shows the average
joint cooperation rate across supergames, focusing on both period-1 cooperation (left panel) and cooperation levels after period 1 (right panel). Two features stand out. First,  cooperation is  considerably higher in the presence of communication. In the HAI treatments, cooperation without the chat option is 45.3\% on average and rises to 81.2\% with pre-play communication, and reaches 83.2\% with repeated communication.
Second, both communication protocols achieve similar cooperation levels (p= 0.699), so almost all of the communication effect is delivered by a single exchange before the supergame begins.

Comparing HAI-Rep and HH-Rep which differ only in the partner, cooperation is 14.2 percentage points higher when the partner is an AI (83.2\% against 69.0\%, p<0.001). The gap is present from the first round of each supergame, when neither player has yet observed the other's behaviour.

\begin{figure}[H]
\centering

\begin{subfigure}{0.48\textwidth}
    \centering
    \includegraphics[width=\linewidth]{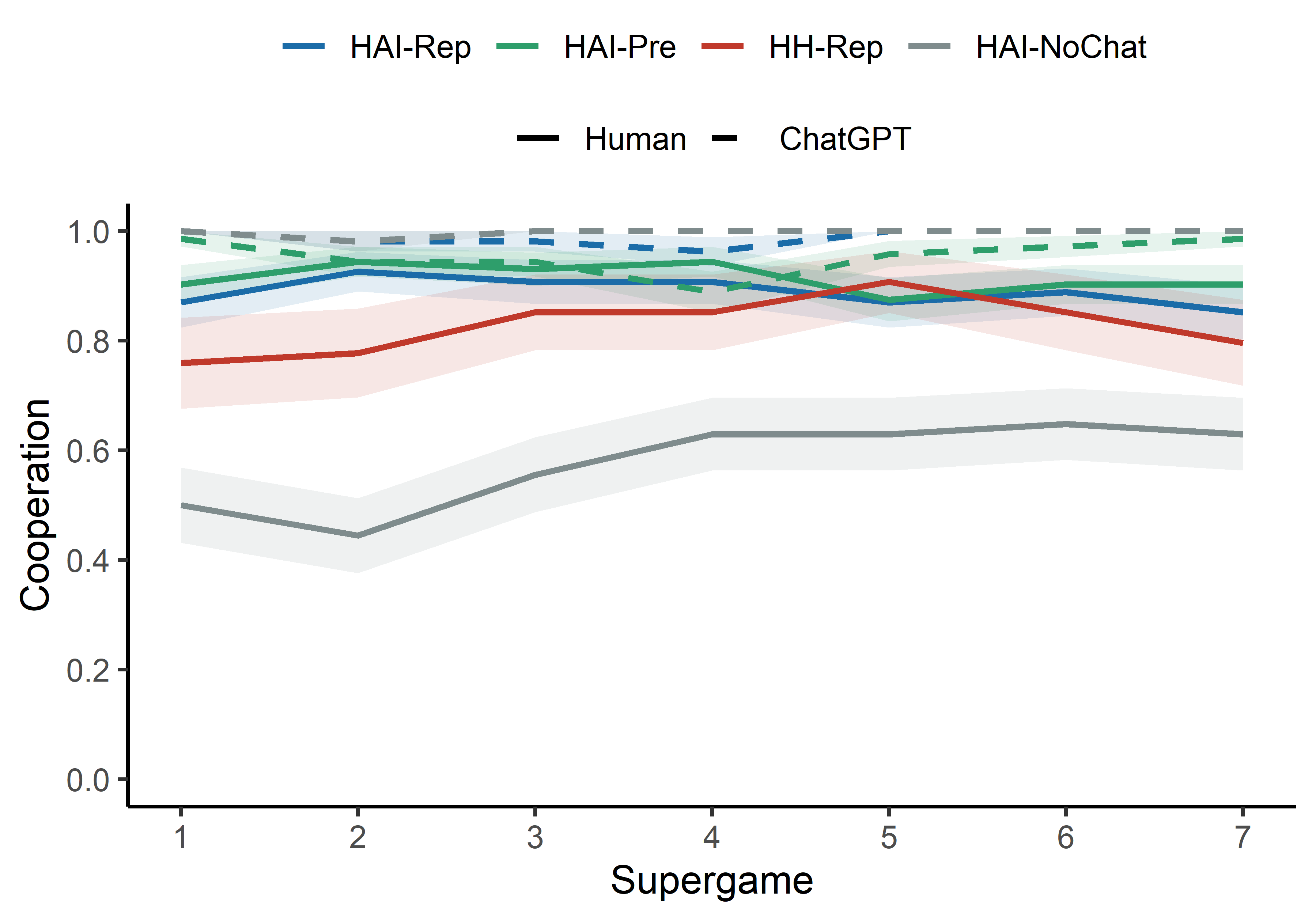}
    \caption{Round 1}
    \label{fig:pre_all}
\end{subfigure}
\hfill
\begin{subfigure}{0.48\textwidth}
    \centering
    \includegraphics[width=\linewidth]{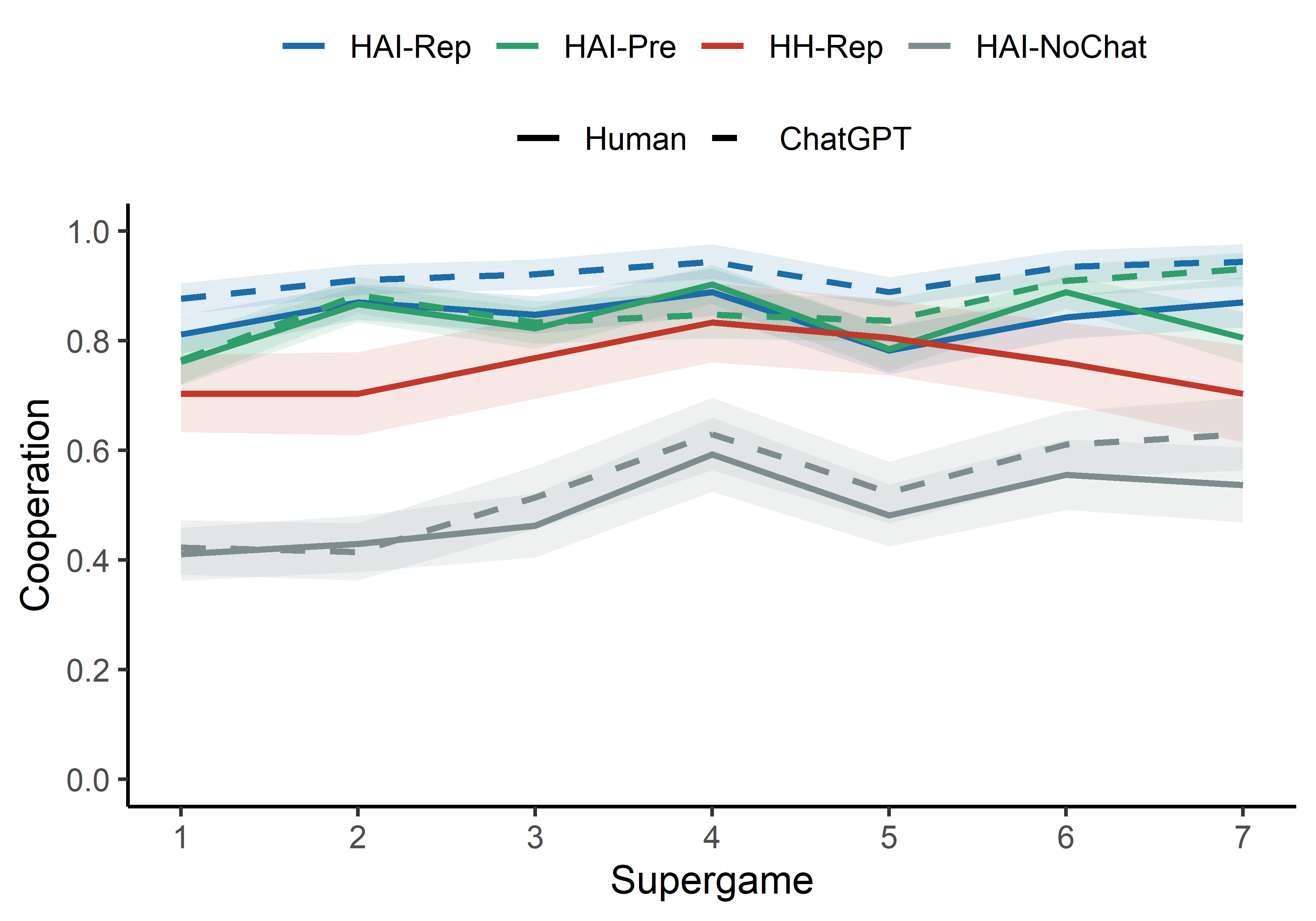}
    \caption{Rounds after round 1}
    \label{fig:rep_all}
\end{subfigure}

\caption{Cooperation by supergame. Left: first round of each supergame,
before either player has observed the other within that supergame. Right:
subsequent rounds. Solid lines show participants' own cooperation, dashed
lines their ChatGPT partner's; HH-Rep has no dashed line as both players are
human. Shaded areas denote $\pm 1$ standard error of the mean.}
\label{fig:coop_split}
\end{figure}

Figure~\ref{fig:coop_split} decomposes the HAI treatments into the human's own cooperation and their AI partner's cooperation by supergame. The left panel shows decisions only in round 1 while the right panel shows all subsequent rounds. In the first round of each supergame, ChatGPT cooperates in essentially every supergame of every treatment (96--100\% in HAI-Rep, 89--99\% in HAI-Pre, 98--100\% in HAI-NoChat), while the human's opening rate ranges from 50--65\% in
HAI-NoChat to 85--93\% in HAI-Rep. The AI's opening play is therefore invariant across treatments, and the entire treatment effect on first-round behaviour can be attributed to human players.

Table~\ref{tab:cooperation} summarises these patterns, reporting mean joint cooperation for the early, late, and full sets of supergames together with pairwise statistical tests. These tests are two-sided Mann-Whitney rank-sum tests with one observation per independent unit: the individual human subject in the HAI treatments and the interacting pair in HH-Rep
\begin{fnd}
Communication substantially raises cooperation against an AI partner, essentially all of it delivered by a single pre-play exchange; allowing communication within rounds produces no measurable further gain. 
\end{fnd}

\begin{table}[htbp]
\centering
\footnotesize
\caption{Joint cooperation by treatment}
\label{tab:cooperation}
\begin{tabular}{lccc}
\toprule
 & Supergames 1--3 & Supergames 5--7 & All supergames \\
\midrule
\multicolumn{4}{l}{\emph{Panel A: mean joint cooperation}} \\
\addlinespace
HAI-Rep ($n=54$) & 0.826 & 0.823 & 0.832 \\
 & (0.261) & (0.259) & (0.232) \\
HAI-Pre ($n=72$) & 0.791 & 0.826 & 0.812 \\
 & (0.251) & (0.291) & (0.232) \\
HH-Rep ($n=54$) & 0.651 & 0.699 & 0.690 \\
 & (0.333) & (0.351) & (0.264) \\
HAI-NoChat ($n=54$) & 0.343 & 0.522 & 0.453 \\
 & (0.352) & (0.422) & (0.361) \\
\addlinespace
\midrule
\multicolumn{4}{l}{\emph{Panel B: pairwise comparisons}} \\
\addlinespace
HAI-Rep vs.\ HAI-Pre & 0.639 & 0.815 & 0.699 \\
 & [2035.0, 0.45] & [1901.5, $-$0.21] & [2020.5, 0.38] \\
HAI-Rep vs.\ HH-Rep & 0.003 & 0.075 & $<$0.001 \\
 & [1931.0, 2.91] & [1726.0, 1.65] & [2033.5, 3.54] \\
HAI-Rep vs.\ HAI-NoChat & $<$0.001 & $<$0.001 & $<$0.001 \\
 & [2430.5, 5.98] & [2059.0, 3.69] & [2350.0, 5.48] \\
HAI-Pre vs.\ HH-Rep & 0.012 & 0.037 & 0.002 \\
 & [2437.5, 2.43] & [2331.0, 1.91] & [2575.5, 3.11] \\
HAI-Pre vs.\ HAI-NoChat & $<$0.001 & $<$0.001 & $<$0.001 \\
 & [3235.0, 6.36] & [2746.5, 3.96] & [3077.0, 5.59] \\
HH-Rep vs.\ HAI-NoChat & $<$0.001 & 0.029 & 0.001 \\
 & [2124.5, 4.10] & [1802.5, 2.12] & [1982.5, 3.22] \\
\bottomrule
\end{tabular}

\vspace{0.5em}
\begin{minipage}{0.9\textwidth}
\footnotesize
\emph{Notes:} Joint cooperation is the share of rounds in which both members of a pair cooperated. Panel~A reports means across subject--supergame observations, with standard deviations of subject-level means in parentheses. Panel~B reports two-sided Mann--Whitney rank-sum tests: $p$-values on the first line and $[U, z]$ in brackets beneath, where $z$ is the standardised statistic from the normal approximation. The independent observation unit is the individual participant in all treatments, contributing the mean of their per-supergame joint cooperation rates. Supergame lengths were randomly determined, so supergames 5--7 contribute fewer rounds per subject than supergames 1--3.
\end{minipage}
\end{table}

\subsection{Strategy Frequency Estimation Method}
\label{sec:sfem}

In this subsection we use structural econometric modelling to explore whether facing an AI opponent shifts the distribution of strategies humans adopt in the indefinitely repeated PD. Given the discrete nature of our data, we use the
\textit{Strategy Frequency Estimation Method} (SFEM),  introduced in \citet{dal2011} and \citet{fudenberg12} and extensively used since.\footnote{A non-exhaustive list includes \citet{bigoni2015time}, \citet{breitmoser2015cooperation},  \citet{frechette2017infinitely} and
\citet{dal2019}. \citet[p.~3930]{dal2019} provide an overview of alternative
methods and the identification issues that characterise them.}

The method specifies a set of candidate strategies $\mathcal{K}$ and estimates
their frequencies in a finite-mixture model that allows for implementation
error. Subject $i$ plays strategy $k \in \mathcal{K}$ with probability
$\pi_k$, and conditional on $k$ follows the prescribed action in each round
with probability $1-\gamma$, trembling with probability $\gamma$. Letting
$I_{iR}^k$ equal 1 if the action prescribed by strategy $k$ in round $R$
coincides with subject $i$'s observed action, the likelihood of $i$'s sequence
of choices under $k$ is
\[
P_i(s_i \mid k) = \prod_{R=1}^{T_i}
(1-\gamma)^{I_{iR}^k}\,\gamma^{\,1-I_{iR}^k},
\]
and the estimates $(\hat{\gamma}, \hat{\pi})$ maximise
$\sum_{i} \ln \left( \sum_{k} \pi_k P_i(s_i \mid k) \right)$ subject to
$\pi_k \geq 0$ and $\sum_k \pi_k = 1$.

\citet{dalbo2018} find that the strategic behaviour of most subjects can be
classified into a small number of pure strategies, and \citet{heller2023} reach
a similar conclusion using a clustering algorithm. We report estimates for six
strategies: always defect (ALLD), always cooperate (ALLC), Grim Trigger
(GRIM), Tit-for-Tat (TFT), Win-Stay Lose-Shift (WSLS), and T2, a trigger
strategy with two punishment periods.\footnote{\citet{dvorak2024} list and
estimate 24 strategies. We focus on the most commonly observed ones for two
reasons: it allows a clear test of a strategy shift between HAI and HH, and
given the relatively short histories, a large strategy set risks overfitting
and under-identification. We also tried combinations including RAND and STFT,
and report the one with the best fit by the Bayesian Information Criterion.}
We assume that tremble probabilities are partitioned by treatment. Therefore, there are five free mixing probabilities for each treatment, together with the tremble parameter $\gamma$. We estimate the model by maximum likelihood using the R package stratEst \citep{dvorak2023stratest}\footnote{The estimation was conducted using the \emph{R} programming language for statistical computing (The \emph{R} Manuals, version 4.3.1. Available at: http://www.r-project.org/). The data along with the estimation codes are available in the replication folder.}. Table \ref{tab:sfem_humans} reports the SFEM estimates at the treatment level, with all supergames included using only the data from the human players, while Table \ref{tab:sfem_ai} reports the same information on the AI partners. 

From the tables three patterns emerge. First, communication eliminates unconditional defection. ALLD drops from 40.2\% in the HAI-NoChat treatment to essentially zero in all the other treatments. Second,  humans play different conditional strategies against an AI than against a human, even under identical communication. Comparing HAI-Rep to HH-Rep: TFT falls by 39.5 points (p = 0.001), and ALLC gains 25.5 points (p = 0.062). This shows that subjects reciprocate against another human player, but cooperate unconditionally against AI. Finally, the strategic role of communication differs against the AI and the human opponent. In the HAI treatments, communication shifts human players from GRIM (55.6\% under pre-play) to ALLC (40.1\% under repeated), with the GRIM share falling by 43.2 points from pre-play to repeated (p = 0.005), highlighting how repeated communication shifts the strategy distribution against an AI opponent.

The AI's estimated strategy also varies across treatments, nevertheless, this variation is due to the  histories it faced rather than a change in the AI itself. Table~\ref{tab:sfem_ai} shows that, for  AI, TFT is the modal strategy in every treatment (90.3\%, 47.3\%, 64.2\%), with a large GRIM share only  in the pre-play communication (42.9\%). The pattern is not monotone in the communication protocol, and the AI's directly observed behaviour is invariant: it opens with cooperation in at least nine out of ten supergames in every treatment. Since the AI conditions on what its partner does and human behaviour differs sharply across treatments, its estimated strategy moves with the realised histories. We therefore read these estimates as descriptive of observed AI play rather than as evidence that the model adopts a different strategy depending on the communication protocol.
Table \ref{tab:sfem_comparisons} reports the pairwise differences in estimated strategy shares between treatments and reports the significance tests (Wald-tests on estimated shares). The main conclusion can be summarised as differences between treatments are large where communication is introduced and where the partner changes. By contrast, they are small where two treatments differ only in the timing of chat. 

Nevertheless, one has to be careful when interpreting the estimates of the SFEM analysis when the observed histories contain little or no defection. In such cases, several strategically different strategies can become observationally equivalent. For example, when no defection occurs, ALLC, GRIM, and TFT all prescribe cooperation along the realised history, while their predictions differ only following an off-path defection. Thus, when cooperation is very high, the data provide limited information with which to distinguish between unconditional cooperation and different forms of conditional cooperation or punishment. This is particularly relevant for the later supergames in our experiment, which are both short and characterised by high levels of cooperation. Only rounds that follow a defection within the supergame carry information for distinguishing among the candidate strategies, since on any history in which no defection has yet occurred ALLC, GRIM, TFT, WSLS and T2 all prescribe cooperation. The share of such rounds is 46.5\% in HAI-NoChat, 28.5\% in HH-Rep, 16.1\% in HAI-Rep and 15.8\% in HAI-Pre, and falls further in the last three supergames, to 33.3\%, 19.3\%, 14.8\% and 13.1\% respectively. Identification is therefore weakest precisely where cooperation is highest, in the two human--AI communication treatments, where fewer than one round in six is informative. The estimated strategy frequencies should be read as describing observed play that is consistent with a set of strategies, rather than as cleanly identifying the rule a subject would follow off the equilibrium path.

\begin{table}[htbp]
\centering
\footnotesize
\caption{Estimated strategy shares, human participants}
\label{tab:sfem_humans}
\begin{tabular}{lcccc}
\toprule
 & HAI-NoChat & HAI-Pre & HAI-Rep & HH-Rep \\
\midrule
ALLD & 0.402$^{***}$ & 0.024 & 0.055 & 0.053 \\
 & (0.069) & (0.019) & (0.031) & (0.033) \\
ALLC & 0.056 & 0.137$^{*}$ & 0.401$^{***}$ & 0.146$^{*}$ \\
 & (0.047) & (0.070) & (0.119) & (0.067) \\
TFT & 0.094 & 0.081 & 0.000 & 0.395$^{***}$ \\
 & (0.062) & (0.061) & (0.068) & (0.092) \\
GRIM & 0.400$^{***}$ & 0.556$^{***}$ & 0.124 & 0.283$^{***}$ \\
 & (0.083) & (0.111) & (0.107) & (0.081) \\
WSLS & 0.048 & 0.120 & 0.212 & 0.077 \\
 & (0.051) & (0.076) & (0.137) & (0.053) \\
T2 & 0.000$^{***}$ & 0.082 & 0.207 & 0.046 \\
 & (0.000) & (0.090) & (0.160) & (0.042) \\
\midrule
$\gamma$ & 0.165 & 0.090 & 0.100 & 0.092 \\
 & (0.005) & (0.003) & (0.004) & (0.003) \\
Observations & 1,620 & 2,160 & 1,620 & 1,620 \\
\bottomrule
\end{tabular}

\vspace{0.5em}
\begin{minipage}{0.9\textwidth}
\footnotesize
\emph{Notes:} Estimated strategy shares from the Strategy Frequency Estimation Method, pooling all supergames. Standard errors in parentheses. $\gamma$ is the estimated tremble probability. Stars indicate a Wald test of the share against zero:  $^{*}$ $p<0.05$, $^{**}$ $p<0.01$, $^{***}$ $p<0.001$.
\end{minipage}
\end{table}

\begin{table}[htbp]
\centering
\footnotesize
\caption{Estimated strategy shares, AI partner}
\label{tab:sfem_ai}
\begin{tabular}{lccc}
\toprule
 & HAI-NoChat & HAI-Pre & HAI-Rep \\
\midrule
ALLD & 0.000 & 0.000 & 0.000 \\
 & (0.000) & (0.000) & (0.000) \\
ALLC & 0.000$^{*}$ & 0.050 & 0.233$^{**}$ \\
 & (0.000) & (0.035) & (0.084) \\
TFT & 0.903$^{***}$ & 0.473$^{***}$ & 0.642$^{***}$ \\
 & (0.061) & (0.097) & (0.102) \\
GRIM & 0.075 & 0.429$^{***}$ & 0.000$^{***}$ \\
 & (0.058) & (0.098) & (0.000) \\
WSLS & 0.000$^{*}$ & 0.000$^{***}$ & 0.124 \\
 & (0.000) & (0.000) & (0.075) \\
T2 & 0.023 & 0.048 & 0.000$^{***}$ \\
 & (0.022) & (0.044) & (0.000) \\
\midrule
$\gamma$ & 0.030 & 0.024 & 0.034 \\
 & (0.004) & (0.002) & (0.002) \\
Observations & 1,620 & 2,160 & 1,620 \\
\bottomrule
\end{tabular}

\vspace{0.5em}
\begin{minipage}{0.9\textwidth}
\footnotesize
\emph{Notes:} Estimated strategy shares from the Strategy Frequency Estimation Method, pooling all supergames. Standard errors in parentheses. $\gamma$ is the estimated tremble probability. Stars indicate a Wald test of the share against zero: $^{*}$ $p<0.05$, $^{**}$ $p<0.01$, $^{***}$ $p<0.001$.
\end{minipage}
\end{table}
\begin{table}[htbp]
\centering
\footnotesize
\caption{Pairwise differences in strategy shares}
\label{tab:sfem_comparisons}
\begin{tabular}{llccc}
\toprule
Comparison & Strategy & Difference & 95\% CI & $p$ \\
\midrule
HAI-Rep vs HAI-Pre & ALLD & 0.031 & [-0.04, 0.10] & 0.393 \\
 & ALLC & 0.264 & [-0.01, 0.53] & 0.055 \\
 & TFT & -0.081 & [-0.26, 0.10] & 0.376 \\
 & GRIM & -0.432$^{**}$ & [-0.73, -0.13] & 0.005 \\
 & WSLS & 0.092 & [-0.21, 0.40] & 0.557 \\
 & T2 & 0.125 & [-0.23, 0.49] & 0.495 \\
\addlinespace
HAI-Rep vs HH-Rep & ALLD & 0.002 & [-0.09, 0.09] & 0.965 \\
 & ALLC & 0.255 & [-0.01, 0.52] & 0.062 \\
 & TFT & -0.395$^{***}$ & [-0.62, -0.17] & 0.001 \\
 & GRIM & -0.158 & [-0.42, 0.10] & 0.237 \\
 & WSLS & 0.135 & [-0.15, 0.42] & 0.357 \\
 & T2 & 0.161 & [-0.16, 0.49] & 0.330 \\
\addlinespace
HAI-Rep vs HAI-NoChat & ALLD & -0.346$^{***}$ & [-0.49, -0.20] & 0.000 \\
 & ALLC & 0.345$^{**}$ & [0.09, 0.60] & 0.007 \\
 & TFT & -0.094 & [-0.27, 0.09] & 0.309 \\
 & GRIM & -0.276$^{*}$ & [-0.54, -0.01] & 0.042 \\
 & WSLS & 0.164 & [-0.12, 0.45] & 0.261 \\
 & T2 & 0.207 & [-0.11, 0.52] & 0.197 \\
\addlinespace
HAI-Pre vs HH-Rep & ALLD & -0.029 & [-0.10, 0.05] & 0.447 \\
 & ALLC & -0.009 & [-0.20, 0.18] & 0.926 \\
 & TFT & -0.314$^{**}$ & [-0.53, -0.10] & 0.004 \\
 & GRIM & 0.273$^{*}$ & [0.00, 0.54] & 0.047 \\
 & WSLS & 0.043 & [-0.14, 0.22] & 0.640 \\
 & T2 & 0.036 & [-0.16, 0.23] & 0.716 \\
\addlinespace
HAI-Pre vs HAI-NoChat & ALLD & -0.377$^{***}$ & [-0.52, -0.24] & 0.000 \\
 & ALLC & 0.081 & [-0.08, 0.24] & 0.336 \\
 & TFT & -0.013 & [-0.18, 0.16] & 0.884 \\
 & GRIM & 0.156 & [-0.12, 0.43] & 0.262 \\
 & WSLS & 0.072 & [-0.11, 0.25] & 0.429 \\
 & T2 & 0.082 & [-0.09, 0.26] & 0.363 \\
\addlinespace
HH-Rep vs HAI-NoChat & ALLD & -0.348$^{***}$ & [-0.50, -0.20] & 0.000 \\
 & ALLC & 0.090 & [-0.07, 0.25] & 0.274 \\
 & TFT & 0.301$^{**}$ & [0.08, 0.52] & 0.006 \\
 & GRIM & -0.118 & [-0.35, 0.11] & 0.311 \\
 & WSLS & 0.029 & [-0.12, 0.17] & 0.693 \\
 & T2 & 0.046 & [-0.04, 0.13] & 0.279 \\
\bottomrule
\end{tabular}

\vspace{0.5em}
\begin{minipage}{0.9\textwidth}
\footnotesize
\emph{Notes:} Differences in estimated strategy shares between treatments. Standard errors obtained as the square root of the sum of the two variances. $^{*}$ $p<0.05$, $^{**}$ $p<0.01$, $^{***}$ $p<0.001$.
\end{minipage}
\end{table}



\begin{fnd}
The strategy distribution against an AI partner shifts sharply with communication. Without communication, unconditional defection accounts for 40\% of choices; with pre-play or repeated chat, ALLD falls to near zero and the mass moves onto conditional-cooperative strategies (GRIM under pre-play, ALLC under repeated).
\end{fnd}

\begin{fnd}
Under identical communication protocols, humans reciprocate against a human partner but cooperate unconditionally against an AI. 
\end{fnd}

We close this subsection with a comment on who deviates first in the HAI treatments. Cooperation broke down in 64.3\% of HAI-NoChat supergames, 26.8\% of HAI-Pre and 31.7\% of HAI-Rep. Among those, the human defected first in 99.6\%, 81.5\% and 95.0\% of cases respectively, whereas the AI defected first in only 0.4\% of HAI-NoChat breakdowns and in 3--13\% of breakdowns when chat is available. Responses after defection differ between the two players. When chat is not available, the AI punishes almost always and at length (average 2.4 rounds),  and in 59-66\% of the cases it will maintain the punishment until the end of the supergame. Human players are far more forgiving, punishing in 32-72\% of the cases  with an average of 0.5-1.5 rounds. Repeated communication helps to moderate these effects where punishment falls to 0.85 rounds on average  and humans decline to punish at all in 67.9\% of the cases, confirming the repairing function of the repeated communication. 
\subsection{Communication analysis}
\label{sec:comm}
We analyse the free-form communication from the three treatments and ask whether facing an AI opponent changes how humans communicate, and whether what is said predicts what is subsequently done.  Communication content is coded with an explicit, deterministic implementation of the  codebook we develop, validated against independent human coding\footnote{Two coders (the authors) classified 386 speaker-episodes independently; the classifier reproduces the adjudicated human labels on the strategy label in 87.4\% of held-out cases ($\kappa=0.762$), against a HH benchmark of 77.3\% ($\kappa = 0.601$), and the substantive conclusions below are unchanged if the human coding is substituted for the machine coding \citep{BaranskiCooperLee2026}. \citet{BaranskiCooperLee2026} additionally require that the substantive conclusions not depend on who performed the coding. Re-running the analyses reported below on the 386 human-coded observations under the adjudicated human coding rather than     the classifier leaves the prevalence of identifiable rules (33.4\% against 34.7\%), the implementation rate (91.9\% against 90.8\%) and the asymmetry between conditional cooperation and punishment essentially unchanged; all estimated associations with subsequent cooperation retain their sign, and three of four retain significance at the 5\% level. We provide extensive details of our approach in the Online Appendix.}. Analysing text from chat data has long been of interest in experimental economics, and automating that coding is a fast moving topic. For instance \citet{penczynski2019} evaluates simple natural-language and machine-learning tools against previously human-coded transcripts, while a recent strand of the literature explores the role of LLMs in message classification \citep{BaranskiCooperLee2026, CelebiPenczynski2026, FulawkaHertwigWulff2026}. \citet{CelebiPenczynski2026} recommends that effort should be directed at the informational content of the coding instructions rather than at the technology applied to them, and that a codebook should supply what a researcher would need to give a human annotator: experimental context, the underlying theory, category definitions with inclusion and exclusion criteria, and worked examples. Similarly,  \citet{FulawkaHertwigWulff2026} require that each coded category be defined so as to discriminate between choices, rather than capturing a diffuse notion of ``cooperative talk". We design our codebook following these principles and we build a classifier that is a transparent implementation of it\footnote{We classify communication content using a rule-based classifier, an explicit set of lexical and syntactic matching rules derived from the codebook, rather than a supervised or model-based classifier. The advantage of this method is that it is fully transparent and the procedure is exactly reproducible. The classifier is implemented in Python 3.12 \citep{python}. All code and data are available in the replication package.}. We then evaluate the result against the reliability standard of \citet{BaranskiCooperLee2026}, which is agnostic about what model performs the coding\footnote{No classification reported in this paper is model output: every code was produced by the deterministic classifier, which returns identical output on re-execution. An LLM was used to draft the classifier's pattern definitions from the codebook, in the way a research assistant might be asked to operationalise a coding scheme in code; the resulting rules are inspectable and were developed only against the training half of the validation sample. Separately, a model coding of the 286 repeated-communication validation observations under the same codebook agrees with the adjudicated human reference on the strategy label in 85.3\% of cases ($\kappa =0.638$), against a HH benchmark of 83.2\% ($\kappa=0.620$) on those observations. That coding is not used in any reported result}. 

The codebook was written for human annotators, in the tradition of content analysis of
experimental communication and was based on three features. First, the label space is not a generic content taxonomy but the strategy space of the repeated game itself, that is the nine strategy labels are the strategies the theory identifies, so a coded statement is
directly comparable to a behavioural rule. Second, we keep the definitions deliberately conservative and state their exclusions explicitly.\footnote{For instance, ALLC requires an unconditional rule
and is not satisfied by a proposal of joint cooperation or a current-round intention; a statement about the current round alone is NO\_IDENTIFIABLE\_STRATEGY rather than a strategy; and a contingent rule whose response to defection is unstated is preserved as COND\_UNSPECIFIED rather than being forced into TFT or GRIM.} Third, the codebook fixes the unit of observation and the informational boundary around it. The unit of analysis is the speaker-episode: the individual speaker within a communication episode. The coders classified the focal speaker only, use the partner's text solely to interpret meaning, and are blinded to treatment, identity, actions and outcomes.  

First we focus on what is communicated. Table~\ref{tab:content} reports content and stated strategies by treatment. Explicit contingent rules are common in
the human-AI treatments: 59.5\% of pre-play and 41.8\% of repeated statements contain an identifiable repeated-game
rule, and 42.0\% and 28.0\% name a canonical strategy, with TFT accounting for more than 25\% of the messages. On the other hand, Human--human chat is shorter (median two words) and almost entirely rule-free: 3.0\% of statements contain an identifiable rule, and conditional-cooperation, punishment and forgiveness language never appears in the exchange of messages.

\begin{table}[htbp]
\centering
\caption{Communication content and stated strategies, by treatment}
\label{tab:content}
\footnotesize
\begin{tabular}{lrrr}
\toprule
\textbf{} & \textbf{HAI-Pre} & \textbf{HAI-Rep} & \textbf{HH-Rep} \\
\midrule
\multicolumn{4}{l}{\emph{Panel A: communication categories (\% of text-bearing speaker-episodes)}}\\
\quad cooperation proposal & 49.1 & 44.9 & 19.5 \\
\quad commitment promise & 68.3 & 63.1 & 45.2 \\
\quad agreement acceptance & 52.6 & 40.1 & 17.7 \\
\quad conditional cooperation & 45.3 & 30.6 & 0.0 \\
\quad punishment threat & 53.2 & 32.9 & 0.0 \\
\quad forgiveness repair & 12.1 & 10.6 & 0.0 \\
\quad strategy inquiry & 19.4 & 16.3 & 16.7 \\
\quad payoff reasoning & 55.1 & 41.0 & 5.5 \\
\quad belief prediction & 2.8 & 1.8 & 0.1 \\
\quad credibility reliability & 16.9 & 12.7 & 2.3 \\
\quad social affective & 13.2 & 12.8 & 8.9 \\
\quad meta task & 9.2 & 4.7 & 2.0 \\
\addlinespace
\multicolumn{4}{l}{\emph{Panel B: stated strategy (\% of text-bearing speaker-episodes)}}\\
\quad TFT & 27.5 & 26.1 & 0.0 \\
\quad GRIM & 10.4 & 0.9 & 0.0 \\
\quad ALLC & 3.8 & 0.6 & 2.9 \\
\quad ALLD & 0.0 & 0.0 & 0.1 \\
\quad T2 & 0.3 & 0.3 & 0.0 \\
\quad WSLS & 0.0 & 0.1 & 0.0 \\
\quad COND\_UNSPECIFIED & 17.6 & 13.8 & 0.0 \\
\quad NO\_IDENTIFIABLE\_STRATEGY & 40.5 & 58.2 & 97.0 \\
\quad \textit{Any canonical} & 42.0 & 28.0 & 3.0 \\
\quad \textit{Any identifiable rule} & 59.5 & 41.8 & 3.0 \\
\addlinespace
Observations & 974 & 1698 & 932 \\
\bottomrule
\end{tabular}
\begin{minipage}{\linewidth}\vspace{2pt}\scriptsize\emph{Notes.} Speaker-episode level. Both speakers are pooled within HAI episodes.\end{minipage}
\end{table}

We then explore whether the strategies that were explicitly discussed in the chat, were actually implemented during the game. Table~\ref{tab:impl} links each communicated rule to the decision that followed it. For pre-play communication we
evaluate both the first subsequent decision and every round of the subsequent supergame, while for the repeated communication protocol
we use only the immediately following decision, since participants could communicate again and revise their intended
behaviour.\footnote{This is also a validation of the coding itself: following \citet{FulawkaHertwigWulff2026}, who validate identified
content against participants' actual choices, agreement between a communicated rule and the action that follows it is
evidence that the coding captured actual behaviour.} Implementation is high throughout: 95.9\% for tit-for-tat and
97.7\% for unspecified conditional cooperation in pre-play, and 89.6\% and 97.6\% in repeated chat.

Panel D contains the most informative asymmetry. When opponents discuss a conditional cooperative rule, this is followed  almost
always, whereas when the players threaten with  punishment, they do so in only about half of the cases in which their partner
actually defected, providing evidence that promises are kept while threats are largely bluff.

\begin{fnd}
Stated rules are implemented, but asymmetrically: promises to cooperate are honoured in 94--98\% of cases, whereas threats to punish are carried out in only 50\%.
\end{fnd}

\begin{table}[htbp]
\centering
\caption{Implementation: does the communicated rule predict the next action?}
\label{tab:impl}
\footnotesize
\begin{tabular}{lrrr}
\toprule
\textbf{Communicated rule} & \textbf{Statements} & \textbf{Evaluable} & \textbf{Consistent} \\
\midrule
\multicolumn{4}{l}{\emph{Panel A: pre-play, first subsequent decision}}\\
\quad TFT & 268 & 268 & 95.9\% \\
\quad COND\_UNSPECIFIED & 171 & 171 & 97.7\% \\
\quad GRIM & 101 & 101 & 88.1\% \\
\quad ALLC & 37 & 37 & 100.0\% \\
\quad T2 & 3 & 3 & 66.7\% \\
\addlinespace
\multicolumn{4}{l}{\emph{Panel B: pre-play, whole subsequent supergame sequence}}\\
\quad ALLC & 33 & 134 & 85.8\% \\
\quad COND\_UNSPECIFIED & 161 & 658 & 97.3\% \\
\quad GRIM & 83 & 359 & 91.6\% \\
\quad T2 & 3 & 12 & 75.0\% \\
\quad TFT & 253 & 1092 & 95.8\% \\
\addlinespace
\multicolumn{4}{l}{\emph{Panel C: repeated chat, immediate subsequent decision}}\\
\quad TFT & 443 & 443 & 89.6\% \\
\quad COND\_UNSPECIFIED & 234 & 210 & 97.6\% \\
\quad GRIM & 15 & 15 & 86.7\% \\
\quad ALLC & 11 & 11 & 90.9\% \\
\quad T2 & 5 & 1 & 100.0\% \\
\addlinespace
\multicolumn{4}{l}{\emph{Panel D: broader conditional rules, where the condition is observable}}\\
\quad pre-play: conditional cooperation & --- & 441 & 94.3\% \\
\quad repeated: conditional cooperation & --- & 423 & 98.1\% \\
\quad repeated: punishment threat & --- & 108 & 50.0\% \\
\quad repeated: forgiveness / repair & --- & 138 & 49.3\% \\
\bottomrule
\end{tabular}
\begin{minipage}{\linewidth}\vspace{2pt}\scriptsize\emph{Notes.} Panels A and C evaluate the decision immediately following the message. Panel B evaluates every round of the subsequent supergame against the stated rule given the realised history. ``Evaluable'' counts rounds in Panel B and statements elsewhere. A statement is not evaluable when the rule does not pin down an action at that history (e.g.\ COND\_UNSPECIFIED after a partner defection). Rules with fewer than three statements are omitted. Panel D reports the percentage acting as stated among HAI observations where the triggering condition is observed.\end{minipage}
\end{table}

Given that the comparisons above pool both speakers within HAI episodes, and that the AI produces long messages by construction, treatment differences in the pooled data partly reflect who is speaking rather than how humans communicate. We now restrict the analysis using only human speakers. Doing so also clarifies what our design identifies: with three treatments and no HH pre-play condition, the data support two contrasts but not their interaction. HAI-Rep vs HAI-Pre isolates the communication regime, holding the opponent fixed at the AI while HH-Rep vs HAI-Rep isolates the identity of the opponent, holding the regime fixed at repeated.

Table \ref{tab:humanreg} reports both. Columns 2--4 report the percentage of each category while columns 5 and 6 report coefficients from separate logistic regressions, one per category, with dependent variable being an indicator for whether the human speaker's message was coded in that category on two treatment dummies with HAI-Pre as the benchmark. Coefficients are expressed as log-odds, so negative values indicate that the category is less likely to appear compared to the benchmark. Both contrasts point the same way. Switching from pre-play to repeated communication the human messages become less deliberative, that is agreement and payoff reasoning fall sharply, and explicit rule statements collapse from 20.9\% to 3.1\%. Pre-play chat has to specify a plan for a whole supergame, whereas repeated chat allows participants to adjust round by round. Switching the opponent from the AI to another human produces larger effects on every category except commitment (though insignificantly). HH messages may be shorter but still they are able to express the intention of the player, allowing the chat to function as a coordination device rather than a means to communicate strategies.  A potential explanation for the differences observed with respect to the opponent is that humans may talk more strategically to the AI partly due to the way the AI communicates using long strategic messages rather than evidence that people reason harder against machines.

We conclude this section with a note on the link between communicated strategies and the strategies estimated from choices in Section \ref{sec:sfem}. Communicated and estimated strategies are distinct constructs and we do not equate them. Where both are abundant they agree: the AI states a reciprocal or trigger rule in 98\% of the relevant statements, and the SFEM estimate in pre-play on TFT and GRIM is around 90\%. For human participants the two diverge. The SFEM puts more weight to trigger strategies than the chat states (GRIM 0.56 against 0.25 in HAI-Pre), and in HH interaction it attributes 0.40 to TFT when no message states a TFT kind of rule. Nevertheless, one needs to be careful on how to interpret this result given that only 20.9\% of human pre-play statements articulate any rule at all, while the SFEM estimates cover everyone. 
\begin{table}[htbp]
\centering
\caption{Human speakers only: communication content by treatment}
\label{tab:humanreg}
\footnotesize
\resizebox{\textwidth}{!}{%
\begin{tabular}{lrrrcc}
\toprule
\textbf{Category} & \textbf{HAI-Pre} & \textbf{HAI-Rep} & \textbf{HH-Rep} & \textbf{Regime} & \textbf{Opponent} \\
\midrule
cooperation proposal    & 45.8 & 35.2 & 19.5 & $-$0.44 & $-$1.25$^{***}$ \\
commitment promise      & 39.0 & 27.1 & 45.2 & $-$0.54$^{*}$ & $+$0.25 \\
agreement acceptance    & 38.2 & 18.8 & 17.7 & $-$0.98$^{***}$ & $-$1.06$^{***}$ \\
conditional cooperation &  4.7 &  1.2 &  0.0 & n.e. & n.e. \\
punishment threat       & 10.9 &  1.6 &  0.0 & n.e. & n.e. \\
forgiveness repair      &  4.7 &  2.5 &  0.0 & n.e. & n.e. \\
strategy inquiry        & 34.1 & 30.2 & 16.7 & $-$0.18 & $-$0.94$^{***}$ \\
payoff reasoning        & 32.4 & 13.8 &  5.5 & $-$1.10$^{***}$ & $-$2.12$^{***}$ \\
belief prediction       &  2.9 &  1.1 &  0.1 & n.e. & n.e. \\
credibility reliability & 15.4 & 12.1 &  2.3 & $-$0.28 & $-$2.07$^{***}$ \\
social affective        & 17.7 & 19.1 &  8.9 & $+$0.09 & $-$0.79 \\
meta task               &  9.9 &  6.6 &  2.0 & $-$0.44 & $-$1.66$^{**}$ \\
\addlinespace
\textit{any identifiable rule} & 20.9 & 3.1 & 3.0 & --- & --- \\
\bottomrule
\end{tabular}
}
\begin{minipage}{\linewidth}\vspace{2pt}\scriptsize\emph{Notes.} Columns 2–4 report the percentage of human speaker-episodes classified in that category (N=487, 849 and 932). Columns 5–6 report coefficients from separate logit regressions on two treatment dummies with HAI-Pre as the benchmark, with standard errors clustered at the participant level (95, 53 and 66 clusters). "n.e." denotes not estimable. $^{*}p<0.05$, $^{**}p<0.01$, $^{***}p<0.001$.\end{minipage}
\end{table}

\subsection{Regression analysis}
In Table \ref{tab:regressions} we report round-level linear probability models where we use individual cooperation choice as the dependent variable with HAI-NoChat, as the baseline. Since the three treatments vary one dimension at a time rather than
forming a fully crossed design, each coefficient has a direct interpretation:
HAI-NoChat isolates the effect of removing communication against AI as a fixed opponent, while  HH-Rep isolates the effect of replacing the AI with a human while holding communication fixed.\footnote{We note that our
design does not include a human-human treatment without communication, so we
cannot estimate whether the effect of communication itself differs across
opponent types; the comparison between HH-Rep and HAI-NoChat varies both
dimensions simultaneously and is not interpreted.}

The results highlight the importance of 
communication. In the HAI treatments, lack of communication reduces cooperation by 35.8 percentage points ($\beta = -0.358$, $p < 0.001$), an effect that is present from the first round of each supergame ($-0.312$, $p < 0.001$) and persists in the final three
supergames ($-0.281$, $p < 0.001$). Communication affects cooperation in a manner similar to that observed in human--human interactions as  reported by \citet{dvorak2024}.

\begin{fnd}
Communication substantially increases cooperation against an AI partner,
raising cooperation by roughly 36 percentage points relative to a
no-communication baseline.
\end{fnd}

Nevertheless, the identity of the partner matters far less. Replacing the AI with a human changes cooperation by 8.4 percentage points ($\beta = -0.084$, $p = 0.051$),
and this difference becomes even smaller (3.0 percentage points) and non-significant in the last three
supergames ($p = 0.562$). We therefore cannot reject equality between the two
environments once communication is available. 

\begin{fnd}
Once communication is available, we cannot reject that cooperation against an
AI equals cooperation against a human.
\end{fnd}

\begin{table}
\centering
\begin{talltblr}[         
caption={Cooperation across treatments \label{tab:regressions}},
note{ }={The dependent variable is an indicator for the participant's own cooperation. HAI-NoChat is the benchmark. Standard errors clustered at the participant level in parentheses. Column (1) uses all rounds; (2) the first round of each supergame; (3) subsequent rounds; (4) supergames 5--7; (5) subsequent rounds of supergames 5--7.  * p \num{< 0.05}, ** p \num{< 0.01}, *** p \num{< 0.001}},
]                     
{                     
colspec={Q[]Q[]Q[]Q[]Q[]Q[]},
hline{2}={1-6}{solid, black, 0.05em},
hline{10}={1-6}{solid, black, 0.05em},
hline{1}={1-6}{solid, black, 0.08em},
hline{12}={1-6}{solid, black, 0.08em},
column{2-6}={}{halign=c},
column{1}={}{halign=l},
}                     
& (1) & (2) & (3) & (4) & (5) \\
HAI-Pre & \num{0.352}*** & \num{0.338}*** & \num{0.356}*** & \num{0.292}*** & \num{0.307}*** \\
& (\num{0.048}) & (\num{0.054}) & (\num{0.050}) & (\num{0.061}) & (\num{0.065}) \\
HAI-Rep & \num{0.358}*** & \num{0.312}*** & \num{0.372}*** & \num{0.281}*** & \num{0.302}*** \\
& (\num{0.051}) & (\num{0.057}) & (\num{0.053}) & (\num{0.062}) & (\num{0.066}) \\
HH-Rep & \num{0.274}*** & \num{0.251}*** & \num{0.281}*** & \num{0.252}*** & \num{0.267}*** \\
& (\num{0.054}) & (\num{0.061}) & (\num{0.055}) & (\num{0.066}) & (\num{0.069}) \\
Constant (HAI-NoChat) & \num{0.489}*** & \num{0.577}*** & \num{0.462}*** & \num{0.548}*** & \num{0.511}*** \\
& (\num{0.043}) & (\num{0.050}) & (\num{0.043}) & (\num{0.053}) & (\num{0.055}) \\
Observations & \num{7020} & \num{1638} & \num{5382} & \num{2340} & \num{1638} \\
$R^2$ & \num{0.107} & \num{0.114} & \num{0.107} & \num{0.076} & \num{0.080} \\
\end{talltblr}
\end{table}

We close this section by linking our results to those in the literature of indefinitely iterated prisoner's dilemma games with and without communication. The closest study to ours is \citet{dvorak2024}, whose design and parameters (stage-game payoffs and continuation probability) we adopt; the only difference is that signals are noisy in their setting and perfect in ours. Comparing the effects of communication in cooperation, two results are worth mentioning. First, humans cooperate far more with an AI than with a human when there is no communication (0.489 vs. 0.261, $p<0.001$). Secondly, the timing effect of communication (pre-play vs. repeated) observed in \citet{dvorak2024} (+10.9pp, $p=0.023$) does not appear against an AI where the difference is 0.6pp ($p=0.890$). A plausible explanation is that repeated communication helps humans because it lets them renegotiate after defection while the AI agent opens with cooperation in essentially every supergame, reducing the need to renegotiate, and therefore the marginal value of mid-supergame chat collapses\footnote{Effects of this magnitude are common in the wider literature, although these studies differ from ours in payoffs, continuation probability, monitoring structure, or matching protocol, so the rates are not directly comparable to those in Table~\ref{tab:literature}. \citet{andres2023} reports increases of 48 and 57 percentage points from pre-play chat under perfect monitoring, depending on the sucker's payoff, and \citet{boczon2025} find that pre-play chat raises initial cooperation from near zero to almost complete cooperation in a four-player dilemma. \citet{camera2013} on the other hand, find no increase from structured binding promises among strangers under group-level monitoring.}. Note that the noise in \citet{dvorak2024} raises the basin of attraction of defection from 0.274 in our environment to 0.400 in theirs (Table~\ref{tab:literature}, Panel~A). A participant in their setting must therefore hold a stronger belief that the partner will cooperate before cooperation becomes a best response. Lower cooperation is therefore expected in their setting irrespective of the partner's identity, and we do not interpret the difference in levels. 
\begin{table}[htbp]
\centering
\footnotesize
\caption{Cooperation in this paper and in \citet{dvorak2024}}
\label{tab:literature}
\begin{tabular}{lcc}
\toprule
 & This paper & \citet{dvorak2024} \\
\midrule
\multicolumn{3}{l}{\emph{Panel A: design}} \\
\addlinespace
Stage-game payoffs $(R,S,T,P)$      & $(30,0,37,17)$ & $(30,0,37,17)$ \\
Continuation probability $\delta$   & 0.80 & 0.80 \\
Supergames                          & 7 & 7 \\
Signal accuracy                     & 1.00 & 0.90 \\
Expected payoffs given actions      & $(30,0,37,17)$ & $(27,3,35,19)$ \\
Basin of attraction of defection    & 0.274 & 0.400 \\
Partner                             & AI / human & human \\
\addlinespace
\midrule
\multicolumn{3}{l}{\emph{Panel B: cooperation rate, all rounds}} \\
\addlinespace
No communication                    & 0.489 & 0.261 \\
Pre-play communication              & 0.841 & 0.788 \\
Repeated communication              & 0.847 & 0.897 \\
Repeated communication, human partner & 0.763 & --- \\
\addlinespace
\midrule
\multicolumn{3}{l}{\emph{Panel C: within-study differences}} \\
\addlinespace
Pre-play vs.\ no communication      & $+$0.352$^{***}$ & $+$0.527$^{***}$ \\
Repeated vs.\ pre-play              & $+$0.006         & $+$0.108$^{*}$ \\
AI vs.\ human partner, repeated     & $+$0.084   & --- \\
\bottomrule
\end{tabular}

\vspace{0.5em}
\begin{minipage}{0.95\textwidth}
\footnotesize
\emph{Notes:} Entries in Panels~B and~C are the rate at which the participant
cooperates, pooled over all rounds. The \citet{dvorak2024} column is computed
by us from their replication data, restricted to treatments in which both
players observe the same signal. Differences are in percentage points, with
$p$-values from two-sided Mann--Whitney tests on subject-level means.
The two studies share stage-game payoffs, $\delta$, and the number of
supergames, but in \citet{dvorak2024} actions are translated into signals that
are correct with probability $0.9$ and payoffs follow the signal rather than
the action. The expected payoffs conditional on actions therefore differ, and
the basin of attraction of defection is larger in their environment, so lower
cooperation is expected there irrespective of the partner's identity.
Levels are not comparable across columns. 
$^{*}$ $p<0.05$, $^{**}$ $p<0.01$, $^{***}$ $p<0.001$.
\end{minipage}
\end{table}
\section{Conclusion}
\label{sec:conclusion}

This paper provides controlled evidence on how natural language communication with an AI partner affects human cooperation in indefinitely repeated interaction. By varying communication against a chatbot powered by a large language model and, under the every-round protocol, replacing the chatbot with another human, the design identifies both the effect of communication and the effect of the partner's identity within a single experiment.

Four findings emerge. First, communication is what sustains cooperation with the AI, and a single exchange is enough. One pre-play exchange raises cooperation from 49\% to 84\%, and chatting before every round adds less than one percentage point, whereas between humans repeated communication raises cooperation further \citep{dvorak2024}. Second, once communication is available, cooperation with the AI is no lower than with a human partner. Third, the partner's identity shifts strategies. Under the identical every-round protocol, Tit-for-Tat is the modal strategy against a human and unconditional cooperation the modal strategy against the AI. Fourth, what participants say predicts what they do, although not symmetrically. Promises to cooperate are almost always kept, whereas threats to punish are carried out only about half the time.

These results suggest that communication with an AI works through a single channel, coordination on the cooperative equilibrium. The AI follows the rule it states and rarely defects first, so there is little for ongoing conversation to repair, which is why the timing of communication does not matter. The asymmetry between promises and threats is consistent with the view in social psychology that discussion raises cooperation because it elicits commitments that people feel bound to honour \citep{kerr1994}, and our data indicate that this holds when the promise is made to a machine. The results also qualify the ``machine penalty'' in cooperative games, which has been reported to persist over repeated interaction \citep{bonnefon2024}. The cross-study comparison is consistent with the machine penalty, as participants frequently defect on an AI that opens cooperatively, the exploitation of a benevolent machine documented by \citet{karpus2021}. With communication it disappears. One reading, in the terms of \citet{qin2025}, is that a partner who announces a simple rule and keeps to it is perceived as capable in a context where personalised treatment is unnecessary, which are the conditions under which people respond favourably to AI.

For the design and deployment of AI agents, the implication is that cooperation did not require anthropomorphic cues, emotional language, or repeated reassurance. It required an opportunity to agree on a rule and an agent whose behaviour matched its words. In settings such as automated negotiation or advice, transparency about the agent's strategy, and consistency in following it, may matter more than conversational richness.

Several limitations should be noted. We use a single model (GPT-5.2) with a fixed system prompt, and different models or prompts may elicit different behaviour on both sides. Cooperation is high in the communication treatments, which leaves few off-path histories and limits how sharply the strategy estimates can separate unconditional from conditional cooperation. Finally, our content analysis captures only the rules that participants articulate, and most human statements contain no identifiable rule.

The treatment differences originate on the human side, since the AI behaves in much the same way in every condition. What we cannot say is why. A natural next step is to elicit beliefs about how an AI partner will play, and to test how far behaviour is driven by those beliefs, which may be inaccurate, rather than by the AI's actual decisions. Further work could vary the AI's strategy or communication style, use prompts that make the AI more social or emotional, and test whether the patterns documented here extend beyond the Prisoner's Dilemma.

\newpage
\begin{appendices}
\setcounter{table}{0}
\setcounter{figure}{0}
\renewcommand{\thetable}{A\arabic{table}}
\renewcommand{\thefigure}{A\arabic{figure}}
\section{Communication analysis}
\label{sec:commapp}

In this Appendix we provide the details of our communication analysis.  We analyse the free-form communication from the three treatments and ask whether facing an AI opponent changes how
humans communicate, and whether what is said predicts what is subsequently done. Communication content is coded with
an explicit, deterministic implementation of our codebook, validated against independent human coding.  \citet{CelebiPenczynski2026} show that  classification performance is driven primarily by the informational detail of the
codebook rather than by the technology applied to it. Based on this finding, we develop an analytical codebook, and the classifier is a transparent implementation of it. We evaluate the result against
the reliability standard of \citet{BaranskiCooperLee2026}, which is agnostic about who or what model performs the coding.

First, following the standard approach of content analysis, the codebook was written with human annotators in mind, providing the informational content that such an annotator would need: experimental context, the underlying theory, category definitions
with inclusion and exclusion criteria, and worked examples. We also follow \citet{FulawkaHertwigWulff2026} in requiring that each
coded category be defined so as to discriminate between choices, rather than capturing a diffuse notion of ``cooperative talk''. That is, we define the  nine strategy labels as the theory identifies them ,so a coded statement is
directly comparable to a behavioural rule.  Second, the definitions are deliberately conservative and state their exclusions explicitly. For instance, ALLC requires an unconditional rule and is not satisfied by a proposal of joint cooperation or a current-round intention. Such statements are classified as NO\_IDENTIFIABLE\_STRATEGY, whereas  a contingent rule whose response to defection
is unstated is coded as COND\_UNSPECIFIED rather than being forced into TFT or GRIM. Third, the codebook fixes the unit of observation and the informational boundary around it. Coders classify the focal speaker only, use the partner's text solely to interpret meaning, and are blinded to treatment, identity,
actions and outcomes.

To assess whether  the codebook is well specified we tested  convergence between two independent classifications. Two human
coders (the authors), working from it independently, agreed on the strategy label in 77--83\% of cases, and a mechanical implementation of the same
document reproduced the adjudicated human labels at $\kappa = 0.762$ on held-out episodes not used in its development.
Agreement of that kind is a property of the coding scheme rather than of any particular coder, and it is the coding scheme that 
\citet{CelebiPenczynski2026} identify as the binding constraint on classification quality.

 The codebook distinguishes twelve binary communication categories and nine mutually exclusive strategy labels (ALLC, ALLD, TFT,
GRIM, T2, WSLS, COND\_UNSPECIFIED, OTHER, NO\_IDENTIFIABLE\_STRATEGY). TFT is applied when the speaker conditions
cooperation on the partner's cooperation and specifies a one-period-lagged response to defection (``if you play B,
I'll switch to B next round''); GRIM requires punishment to be explicitly permanent.
\subsection{The classifier}
We classify communication content using a rule-based classifier, an explicit set of lexical and syntactic matching rules derived from the codebook, rather than a supervised or model-based classifier. This choice is deliberate: the method is fully transparent, every classification traces to an identifiable span of text, and the procedure is exactly reproducible\footnote{The advantage of this method is that it is fully transparent and the procedure is exactly reproducible. The classifier is implemented in Python 3.12 \citep{python}. All code and data are available in the replication package.}.

The codebook is applied by a deterministic program that takes the focal speaker's text and the partner's text from the same episode as its only inputs. It has no access to treatment, participant identity, actions, payoffs or outcomes, which can be verified from the function signature. The implementation is 357 lines of Python, organised into 31 pattern families comprising approximately 1,150 surface variants, and is included in the replication package.

Processing proceeds in three stages. First, text is normalised (Unicode punctuation folded, whitespace collapsed, case lowered) and segmented: an episode is split into messages, and each message into clauses at sentence punctuation and coordinating conjunctions. Segmentation matters because several categories are defined at clause level rather than message level.

Second, each of the twelve communication categories is evaluated. Most are detected by matching pattern families against clauses, but three require structural conditions that reflect the codebook's exclusions. A \emph{cooperation proposal} requires a hortative or imperative clause directed at joint play (``let's both choose A'', ``stick with A''); a clause that states a payoff consequence conditionally (``if we both pick A we each get 30'') is explicitly excluded, since it reasons about payoffs rather than proposing a course of action. \emph{Conditional cooperation} requires that the speaker's own intended action appear in the same sentence as the condition, and that the condition be on the partner \emph{cooperating} rather than defecting, which separates it from a punishment threat. \emph{Agreement} distinguishes an explicit acceptance from a bare acknowledgement; a bare ``ok'' is coded only when the partner's text contains a plan to accept, which is the one place the codebook permits the partner's text to affect a code.

Third, a single strategy label is assigned by a fixed priority ordering, evaluated on the text with interrogative clauses removed so that questions such as ``will you always choose A?'' cannot be read as an unconditional rule. The order is: WSLS (explicit win-stay/lose-shift language); ALLD and ALLC (an unconditional first-person rule and no conditional or punitive content); T2 (punishment stated as exactly two rounds); GRIM (punishment with an explicit permanence marker --- ``from then on'', ``forever'', ``for the rest of the game'' --- and no one-round qualifier); TFT (explicit copying language, a one-round punishment, or a one-period-lagged response combined with either a punishment threat or a stated return to cooperation); 
COND\_UNSPECIFIED (conditional or punitive content referring to future rounds, without the wording needed to identify a canonical rule); and NO\_IDENTIFIABLE\_STRATEGY otherwise. Priority is strict, so permanence dominates lag: a statement containing both ``I'll switch to B next round'' and ``and stay there'' is GRIM, not TFT.

For each positive category and for the strategy label, the program also returns the text span that triggered it, which is what the human coders reviewed when adjudicating boundary cases. Because the procedure contains no random component and no model inference, re-running it on the same corpus reproduces the classification exactly.

\subsection{Validation}
To validate the codebook, we randomly selected 200 episodes: 150 representative of the corpus and 50 ``challenge'' episodes containing  strategically informative or difficult cases, yielding 386 speaker-episodes. Two coders classified every speaker-episode independently, and  disagreements were adjudicated using only the focal text, the context text and the frozen codebook.

We then split the 200 episodes in half at random, stratified by treatment and by sample component. The coding rules were developed against the first half only and the second half was used once, for the evaluation reported
here (evaluating rules on the same observations used to refine them would overstate their accuracy). We evaluate the result against the two reliability criteria of \citet{BaranskiCooperLee2026}: (C1) the
coding should differ from a human coding by no more than two human codings differ from each other, and (C2) the
qualitative conclusions should not depend on which coding is used.
Table~\ref{tab:crit1} reports C1 on the held-out episodes. Agreement on the strategy label is 87.4\%
($\kappa = 0.762$) against a human--human benchmark of 77.3\% ($\kappa = 0.601$), and C1 is met on ten of the
thirteen coded fields. It is not met for conditional cooperation ($\kappa = 0.814$ against 0.868) or strategy
inquiry (0.695 against 0.722), where the two human coders agree unusually closely with each other, nor for belief
prediction, a rare category on which neither the classifier nor the human coders achieve above-chance agreement and
which we therefore do not interpret. The largest residual error is that six of the eighteen held-out GRIM statements
are coded COND\_UNSPECIFIED, so the incidence of GRIM reported below in \ref{sec:A4}A.5 and \ref{sec:A5} should be read as a lower bound.\footnote{As an additional validation exercise, we asked an LLM to code the 286 repeated-communication observations (the classification was produced by Claude Opus 5 on 15 August 2026, using the frozen codebook verbatim as the prompt and blinded to treatment, identity and outcomes). The classification agrees with the adjudicated human reference on the strategy label in 85.3\% of cases ($\kappa = 0.638$), against a human--human benchmark of
83.2\% ($\kappa = 0.620$) on the same observations. That coding is not used in any reported result.}

Table~\ref{tab:crit2} reports C2, obtained by linking all 386 human-coded observations to the behavioural record and
re-running the central analyses under each coding. Prevalence, implementation rates and the asymmetry between
promises and threats are close under the two codings, and all four estimated associations share the same sign; three
of the four match on significance at the 5\% level and all four at the 10\% level. We therefore apply the codebook
implementation to the full-corpus of 3{,}604 text-bearing speaker-episodes. The classifier is deterministic, so the
full corpus classification is exactly reproducible from the code and data.

\begin{table}[htbp]
\centering
\caption{Criterion 1: agreement on the held-out validation episodes}
\label{tab:crit1}
\footnotesize
\begin{tabular}{lrrrrc}\toprule
 & \multicolumn{2}{c}{Classifier vs. human reference}& \multicolumn{2}{c}{Human vs. human (pre-adjudication)}&\\
\midrule
\textbf{Field} & \textbf{Agreement} & \textbf{$\kappa$} & \textbf{Agreement } & \textbf{$\kappa$ } & \textbf{C1} \\ \midrule

cooperation proposal & 85.9 & 0.700 & 69.7 & 0.380 & yes \\
commitment promise & 90.9 & 0.809 & 84.3 & 0.661 & yes \\
agreement acceptance & 87.4 & 0.725 & 82.8 & 0.617 & yes \\
conditional cooperation & 92.4 & 0.814 & 94.4 & 0.868 & no \\
punishment threat & 98.0 & 0.952 & 94.4 & 0.863 & yes \\
forgiveness repair & 95.5 & 0.783 & 93.4 & 0.664 & yes \\
strategy inquiry & 91.9 & 0.695 & 91.9 & 0.722 & no \\
payoff reasoning & 90.9 & 0.785 & 80.8 & 0.447 & yes \\
belief prediction & 97.5 & -0.008 & 98.5 & -0.007 & no \\
credibility reliability & 93.4 & 0.687 & 90.9 & 0.463 & yes \\
social affective & 93.4 & 0.749 & 87.9 & 0.463 & yes \\
meta task & 93.4 & 0.550 & 90.9 & 0.393 & yes \\
stated strategy & 87.4 & 0.762 & 77.3 & 0.601 & yes \\
\bottomrule
\end{tabular}
\begin{minipage}{\linewidth}\vspace{2pt}\scriptsize\emph{Notes.} Held-out half of the validation episodes (198 speaker-episodes from 100 episodes), never used while developing the coding rules. Columns 2--3 compare the classifier with the adjudicated human reference; columns 4--5 compare the two human coders with each other pre-adjudication. Criterion 1 of \citet{BaranskiCooperLee2026} is met when the former is no worse than the latter. Belief prediction is a rare category on which neither the classifier nor the two human coders reach above-chance agreement.\end{minipage}
\end{table}

\begin{table}[htbp]
\centering
\caption{Criterion 2: conclusions under human and rule-based coding}
\label{tab:crit2}
\footnotesize
\begin{tabular}{lrr}
\toprule
\textbf{Quantity} & \textbf{Human (adjudicated)} & \textbf{Rule-based} \\
\midrule
Observations & 386 & 386 \\
Participants & 128 & 128 \\
Statements with an identifiable rule \% & 33.4 & 34.7 \\
Statements with a canonical strategy \% & 28.5 & 25.6 \\
Evaluable statements & 124 & 130 \\
Communicated rule implemented \% & 91.9\%& 90.8 \% \\
Cond.\ coop.\ \& partner cooperated $\rightarrow$ A & 96.9\% (n=97) & 94.0 \% (n=84) \\
Threat \& partner defected $\rightarrow$ B & 60.9 \% (n=23) & 47.4\% (n=19) \\
Logit: rule stated $\rightarrow$ cooperation & +0.96 (p=0.000) & +0.95 (p=0.000) \\
Logit: commitment $\rightarrow$ cooperation & +1.06 (p=0.000) & +1.05 (p=0.000) \\
Logit: proposal $\rightarrow$ cooperation & +0.70 (p=0.029) & +0.49 (p=0.092) \\
Logit: cond.\ cooperation $\rightarrow$ cooperation & +0.89 (p=0.000) & +0.55 (p=0.010) \\
\bottomrule
\end{tabular}
\begin{minipage}{\linewidth}\vspace{2pt}\scriptsize\emph{Notes.} All 386 human-coded speaker-episodes (100 pre-play, 286 repeated), linked to the behavioural record. Logits of the immediately subsequent cooperative choice, one predictor at a time, with standard errors clustered at the participant level (128 clusters). Criterion 2 of \citet{BaranskiCooperLee2026} requires that the qualitative conclusions not depend on who did the coding. All signs match; significance matches on three of four associations at the 5\% level and on all four at the 10\% level.\end{minipage}
\end{table}

\subsection{Communication content, full detail}

\begin{table}[H]
\centering
\caption{Prevalence of communication categories, by treatment and stage (\%)}
\label{tab:cats}
\footnotesize
\resizebox{\textwidth}{!}{%
\begin{tabular}{lrrrrrr}
\toprule
\textbf{Category} & \textbf{HAI-Pre} & \textbf{HAI-Rep} & \textbf{HH-Rep} & \textbf{HAI-Pre (ep.)} & \textbf{HAI-Rep (ep.)} & \textbf{HH-Rep (ep.)} \\
\midrule
cooperation proposal & 49.1 & 44.9 & 19.5 & 73.3 & 69.4 & 30.7 \\
commitment promise & 68.3 & 63.1 & 45.2 & 97.5 & 99.2 & 61.3 \\
agreement acceptance & 52.6 & 40.1 & 17.7 & 78.2 & 67.0 & 29.0 \\
conditional cooperation & 45.3 & 30.6 & 0.0 & 85.8 & 60.7 & 0.0 \\
punishment threat & 53.2 & 32.9 & 0.0 & 95.5 & 64.1 & 0.0 \\
forgiveness repair & 12.1 & 10.6 & 0.0 & 20.3 & 18.7 & 0.0 \\
strategy inquiry & 19.4 & 16.3 & 16.7 & 36.8 & 31.6 & 26.2 \\
payoff reasoning & 55.1 & 41.0 & 5.5 & 82.1 & 69.8 & 8.4 \\
belief prediction & 2.8 & 1.8 & 0.1 & 5.5 & 3.5 & 0.2 \\
credibility reliability & 16.9 & 12.7 & 2.3 & 26.1 & 19.9 & 3.7 \\
social affective & 13.2 & 12.8 & 8.9 & 19.3 & 20.5 & 14.1 \\
meta task & 9.2 & 4.7 & 2.0 & 13.1 & 7.5 & 3.3 \\
\bottomrule
\end{tabular}
}
\begin{minipage}{\linewidth}\vspace{2pt}\scriptsize\emph{Notes.} Left panel: speaker-observation level (N = 974 / 1,698 / 932). Right panel: episode level, where an episode counts once if either speaker used the category (N = 487 / 849 / 538). Episode-level figures pool both speakers, including the AI.\end{minipage}
\end{table}

\begin{table}[H]
\centering
\caption{Communication categories by speaker type (\% of speaker observations )}
\label{tab:catsspk}
\footnotesize
\resizebox{\textwidth}{!}{%
\begin{tabular}{lrrrrr}
\toprule
\textbf{Category} & \textbf{HAI-Pre AI} & \textbf{HAI-Pre Human} & \textbf{HAI-Rep AI} & \textbf{HAI-Rep Human} & \textbf{HH-Rep Human} \\
\midrule
cooperation proposal & 52.4 & 45.8 & 54.5 & 35.2 & 19.5 \\
commitment promise & 97.5 & 39.0 & 99.2 & 27.1 & 45.2 \\
agreement acceptance & 66.9 & 38.2 & 61.4 & 18.8 & 17.7 \\
conditional cooperation & 85.8 & 4.7 & 60.1 & 1.2 & 0.0 \\
punishment threat & 95.5 & 10.9 & 64.1 & 1.6 & 0.0 \\
forgiveness repair & 19.5 & 4.7 & 18.7 & 2.5 & 0.0 \\
strategy inquiry & 4.7 & 34.1 & 2.4 & 30.2 & 16.7 \\
payoff reasoning & 77.8 & 32.4 & 68.3 & 13.8 & 5.5 \\
belief prediction & 2.7 & 2.9 & 2.5 & 1.1 & 0.1 \\
credibility reliability & 18.5 & 15.4 & 13.2 & 12.1 & 2.3 \\
social affective & 8.8 & 17.7 & 6.5 & 19.1 & 8.9 \\
meta task & 8.6 & 9.9 & 2.7 & 6.6 & 2.0 \\
\bottomrule
\end{tabular}
}
\begin{minipage}{\linewidth}\vspace{2pt}\scriptsize\emph{Notes.} N = 487 per HAI-Pre column, 849 per HAI-Rep column, 932 for HH-Rep.\end{minipage}
\end{table}

\subsection{Stated strategies by speaker type}\label{sec:A4}

\begin{table}[H]
\centering
\caption{Explicitly stated strategies by speaker type}
\label{tab:stratspk}
\footnotesize
\resizebox{\textwidth}{!}{%
\begin{tabular}{lrrrrr}
\toprule
\textbf{Stated strategy} & \textbf{HAI-Pre AI} & \textbf{HAI-Pre Human} & \textbf{HAI-Rep AI} & \textbf{HAI-Rep Human} & \textbf{HH-Rep Human} \\
\midrule
ALLC & 0.6 & 7.0 & 0.6 & 0.7 & 2.9 \\
ALLD & 0.0 & 0.0 & 0.0 & 0.0 & 0.1 \\
TFT & 51.3 & 3.7 & 51.4 & 0.8 & 0.0 \\
GRIM & 17.2 & 3.5 & 1.6 & 0.1 & 0.0 \\
T2 & 0.6 & 0.0 & 0.6 & 0.0 & 0.0 \\
WSLS & 0.0 & 0.0 & 0.1 & 0.0 & 0.0 \\
COND UNSPECIFIED & 28.3 & 6.8 & 26.1 & 1.4 & 0.0 \\
OTHER & 0.0 & 0.0 & 0.0 & 0.0 & 0.0 \\
NO IDENTIFIABLE STRATEGY & 1.8 & 79.1 & 19.6 & 96.9 & 97.0 \\
ANY canonical (ALLC..WSLS) & 69.8 & 14.2 & 54.3 & 1.6 & 3.0 \\
ANY identifiable rule & 98.2 & 20.9 & 80.4 & 3.1 & 3.0 \\
N & 487.0 & 487.0 & 849.0 & 849.0 & 932.0 \\
\bottomrule
\end{tabular}
}
\end{table}

\subsection{Broader strategic rules discussed}\label{sec:A5}

\begin{table}[H]
\centering
\caption{Broader strategic rules discussed (\% of text-bearing speaker observations)}
\label{tab:broad}
\footnotesize
\resizebox{\textwidth}{!}{%
\begin{tabular}{lrrr}
\toprule
\textbf{Rule type} & \textbf{HAI-Pre} & \textbf{HAI-Rep} & \textbf{HH-Rep} \\
\midrule
unconditional cooperation (ALLC) & 3.8 & 0.6 & 2.9 \\
unconditional defection (ALLD) & 0.0 & 0.0 & 0.1 \\
reciprocal / copying rule (TFT, mirror) & 27.5 & 26.1 & 0.0 \\
permanent trigger (GRIM) & 10.4 & 0.9 & 0.0 \\
fixed-length punishment (T2) & 0.3 & 0.3 & 0.0 \\
conditional cooperation (any, category) & 45.3 & 30.6 & 0.0 \\
punishment after defection (any, category) & 53.2 & 32.9 & 0.0 \\
forgiveness / repair (any, category) & 12.1 & 10.6 & 0.0 \\
cond. coop AND punishment together & 42.2 & 21.6 & 0.0 \\
cooperation proposal, no own rule & 16.6 & 23.9 & 17.3 \\
own commitment, no repeated rule & 13.9 & 22.1 & 45.1 \\
N & 974.0 & 1698.0 & 932.0 \\
\bottomrule
\end{tabular}
}
\begin{minipage}{\linewidth}\vspace{2pt}\scriptsize\emph{Notes.} The final two rows capture statements that propose cooperation or announce a current-round choice without committing to any contingent rule.\end{minipage}
\end{table}

\subsection{Implementation, full detail}

\begin{table}[H]
\centering
\caption{Pre-play implementation: first subsequent decision (round 1)}
\label{tab:preimm}
\footnotesize
\begin{tabular}{lrrrrr}
\toprule
\textbf{Communicated rule} & \textbf{Statements} & \textbf{Evaluable} & \textbf{Implemented} & \textbf{Rate} & \textbf{Not eval.} \\
\midrule
TFT & 268 & 268 & 257 & 95.9\% & 0 \\
COND\_UNSPECIFIED & 171 & 171 & 167 & 97.7\% & 0 \\
GRIM & 101 & 101 & 89 & 88.1\% & 0 \\
ALLC & 37 & 37 & 37 & 100.0\% & 0 \\
T2 & 3 & 3 & 2 & 66.7\% & 0 \\
\bottomrule
\end{tabular}
\begin{minipage}{\linewidth}\vspace{2pt}\scriptsize\emph{Notes.} T2 estimates are descriptive only (N $\leq$ 6).\end{minipage}
\end{table}

\begin{table}[H]
\centering
\caption{Pre-play implementation: whole subsequent supergame sequence}
\label{tab:preseq}
\footnotesize
\resizebox{\textwidth}{!}{%
\begin{tabular}{lrrrrrr}
\toprule
\textbf{Communicated rule} & \textbf{Statements} & \textbf{Rounds eval.} & \textbf{Rounds consist.} & \textbf{\% rounds} & \textbf{SGs fully consist.} & \textbf{ \% SGs fully consist.} \\
\midrule
ALLC & 33 & 134 & 115 & 85.8 & 25.0 & 75.8 \\
COND\_UNSPECIFIED & 161 & 658 & 640 & 97.3 & 151.0 & 93.8 \\
GRIM & 83 & 359 & 329 & 91.6 & 66.0 & 79.5 \\
T2 & 3 & 12 & 9 & 75.0 & 2.0 & 66.7 \\
TFT & 253 & 1092 & 1046 & 95.8 & 227.0 & 89.7 \\
\bottomrule
\end{tabular}
}
\begin{minipage}{\linewidth}\vspace{2pt}\scriptsize\emph{Notes.} Each round of the supergame is evaluated against the communicated rule given the history actually realised. AI action paths are reconstructed from the human partner's recorded other.choice, which is an identity rather than an imputation. Coverage: Coverage: 533 of the 580 pre-play statements with an identifiable rule.\end{minipage}
\end{table}

\begin{table}[H]
\centering
\caption{Repeated-chat implementation: immediate subsequent decision (HAI-Rep)}
\label{tab:repimm}
\footnotesize
\begin{tabular}{lrrrrr}
\toprule
\textbf{Communicated rule} & \textbf{Statements} & \textbf{Evaluable} & \textbf{Implemented} & \textbf{Rate} & \textbf{Not eval.} \\
\midrule
TFT & 443 & 443 & 397 & 89.6\% & 0 \\
COND\_UNSPECIFIED & 234 & 210 & 205 & 97.6\% & 24 \\
GRIM & 15 & 15 & 13 & 86.7\% & 0 \\
ALLC & 11 & 11 & 10 & 90.9\% & 0 \\
T2 & 5 & 1 & 1 & 100.0\% & 4 \\
WSLS & 1 & 1 & 1 & 100.0\% & 0 \\
\bottomrule
\end{tabular}
\begin{minipage}{\linewidth}\vspace{2pt}\scriptsize\emph{Notes.} 
A statement is not evaluable when the stated rule prescribes no verifiable action at the history that actually occurred. In practice this arises only after a partner defection, and only for two rules: COND\_UNSPECIFIED, where the speaker announced a response but not its duration, and T2, where the stated two rounds of punishment cannot be verified from a single subsequent decision. In HH-Rep, 28 statements carried an identifiable rule (27 ALLC, 1 ALLD), of which 27 were implemented.\end{minipage}
\end{table}

\begin{table}[H]
\centering
\caption{Repeated-chat implementation by speaker type (HAI-Rep)}
\label{tab:repimmspk}
\footnotesize
\begin{tabular}{lrrrrrr}
\toprule
\textbf{Speaker} & \textbf{Rule} & \textbf{Statements} & \textbf{Evaluable} & \textbf{Implemented} & \textbf{Rate} & \textbf{Not eval.} \\
\midrule
AI & TFT & 436 & 436 & 390 & 89.4\% & 0 \\
AI & COND\_UNSPECIFIED & 222 & 198 & 196 & 99.0\% & 24 \\
AI & GRIM & 14 & 14 & 12 & 85.7\% & 0 \\
AI & ALLC & 5 & 5 & 5 & 100.0\% & 0 \\
AI & T2 & 5 & 1 & 1 & 100.0\% & 4 \\
AI & WSLS & 1 & 1 & 1 & 100.0\% & 0 \\
Human & COND\_UNSPECIFIED & 12 & 12 & 9 & 75.0\% & 0 \\
Human & TFT & 7 & 7 & 7 & 100.0\% & 0 \\
Human & ALLC & 6 & 6 & 5 & 83.3\% & 0 \\
Human & GRIM & 1 & 1 & 1 & 100.0\% & 0 \\
\bottomrule
\end{tabular}
\begin{minipage}{\linewidth}\vspace{2pt}\scriptsize\emph{Notes.} Human cells are too small for inference.\end{minipage}
\end{table}

\begin{table}[H]
\centering
\caption{Broader conditional-rule implementation, evaluated only where the condition is observable}
\label{tab:broadimpl}
\footnotesize
\resizebox{\textwidth}{!}{%
\begin{tabular}{lrrrrr}
\toprule
\textbf{Sample} & \textbf{Rule stated} & \textbf{Condition} & \textbf{N} & \textbf{Chose A} & \textbf{Chose B} \\
\midrule
HAI-Pre& conditional cooperation & partner cooperated / first round & 441 & 94.3\% &  \\
HAI-Pre& punishment threat & partner defected last round & 0 &  & n/a \\
HAI-Pre& forgiveness / repair & after a breakdown & 0 & n/a &  \\
HAI-Rep & conditional cooperation & partner cooperated / first round & 423 & 98.1\% &  \\
HAI-Rep & punishment threat & partner defected last round & 108 &  & 50.0\% \\
HAI-Rep & forgiveness / repair & after a breakdown & 138 & 49.3\% &  \\
HH-Rep & conditional cooperation & partner cooperated / first round & 0 & n/a &  \\
HH-Rep & punishment threat & partner defected last round & 0 &  & n/a \\
HH-Rep  & forgiveness / repair & after a breakdown & 0 & n/a &  \\
\bottomrule
\end{tabular}
}
\begin{minipage}{\linewidth}\vspace{2pt}\scriptsize\emph{Notes.} The cooperative branch of a conditional rule is honoured almost always; the punitive branch about half the time. The forgiveness figure is not non-compliance: such statements typically take the form ``I am choosing B this round, but if you play A I will return to A next round".\end{minipage}
\end{table}

\subsection{Communication content and subsequent cooperation}

\begin{longtable}{llrrr}
\caption{Communication content and cooperation in the immediately subsequent decision}\label{tab:contentcoop}\\
\toprule
\textbf{Sample} & \textbf{Feature} & \textbf{N} & \textbf{Coop. rate} & \textbf{Diff. (pp)} \\
\midrule
\endfirsthead
\toprule
\textbf{Sample} & \textbf{Feature} & \textbf{N} & \textbf{Coop. rate} & \textbf{Diff. (pp)} \\
\midrule
\endhead
\bottomrule
\multicolumn{5}{p{0.95\textwidth}}{\scriptsize \emph{Notes.} Difference is versus observations without that feature. Descriptive associations, not causal effects. Features with fewer than 10 observations are omitted.}
\endlastfoot
HAI-Pre & ALL text-bearing observations & 974 & 94.4\% &  \\
HAI-Pre & cooperation proposal & 478 & 94.6\% & $+$0.4 \\
HAI-Pre & commitment promise & 665 & 95.2\% & $+$2.6 \\
HAI-Pre & agreement acceptance & 512 & 94.5\% & $+$0.4 \\
HAI-Pre & conditional cooperation & 441 & 94.3\% & $-$0.0 \\
HAI-Pre & punishment threat & 518 & 94.6\% & $+$0.5 \\
HAI-Pre & forgiveness repair & 118 & 91.5\% & $-$3.2 \\
HAI-Pre & strategy inquiry & 189 & 93.1\% & $-$1.5 \\
HAI-Pre & payoff reasoning & 537 & 93.9\% & $-$1.1 \\
HAI-Pre & belief prediction & 27 & 81.5\% & $-$13.2 \\
HAI-Pre & credibility reliability & 165 & 89.1\% & $-$6.3 \\
HAI-Pre & social affective & 129 & 93.8\% & $-$0.6 \\
HAI-Pre & meta task & 90 & 94.4\% & $+$0.1 \\
HAI-Pre & explicit repeated-game rule stated & 580 & 95.2\% & $+$2.0 \\
\addlinespace
HAI-Rep & ALL text-bearing observations & 1{,}698 & 85.7\% &  \\
HAI-Rep & cooperation proposal & 762 & 87.1\% & $+$2.5 \\
HAI-Rep & commitment promise & 1{,}072 & 88.5\% & $+$7.5 \\
HAI-Rep & agreement acceptance & 681 & 89.0\% & $+$5.4 \\
HAI-Rep & conditional cooperation & 520 & 85.6\% & $-$0.2 \\
HAI-Rep & punishment threat & 558 & 89.1\% & $+$4.9 \\
HAI-Rep & forgiveness repair & 180 & 58.3\% & $-$30.7 \\
HAI-Rep & strategy inquiry & 276 & 87.7\% & $+$2.3 \\
HAI-Rep & payoff reasoning & 697 & 87.1\% & $+$2.3 \\
HAI-Rep & belief prediction & 30 & 76.7\% & $-$9.2 \\
HAI-Rep & credibility reliability & 215 & 74.0\% & $-$13.5 \\
HAI-Rep & social affective & 217 & 82.5\% & $-$3.7 \\
HAI-Rep & meta task & 79 & 77.2\% & $-$8.9 \\
HAI-Rep & explicit repeated-game rule stated & 709 & 88.0\% & $+$3.9 \\
\addlinespace
HH-Rep & ALL text-bearing observations & 932 & 81.0\% &  \\
HH-Rep & cooperation proposal & 182 & 86.8\% & $+$7.2 \\
HH-Rep & commitment promise & 421 & 84.6\% & $+$6.5 \\
HH-Rep & agreement acceptance & 165 & 84.2\% & $+$3.9 \\
HH-Rep & strategy inquiry & 156 & 71.8\% & $-$11.1 \\
HH-Rep & payoff reasoning & 51 & 90.2\% & $+$9.7 \\
HH-Rep & credibility reliability & 21 & 61.9\% & $-$19.5 \\
HH-Rep & social affective & 83 & 81.9\% & $+$1.0 \\
HH-Rep & meta task & 19 & 68.4\% & $-$12.8 \\
HH-Rep & explicit repeated-game rule stated & 28 & 100.0\% & $+$19.6 \\
\end{longtable}

\begin{table}[htbp]
\centering
\caption{Logit of subsequent cooperation on communication content, one predictor at a time}
\label{tab:logits}
\footnotesize
\begin{tabular}{llrrrrr}
\toprule
\textbf{Sample} & \textbf{Predictor} & \textbf{Coef.} & \textbf{SE (clustered)} & \textbf{$p$} & \textbf{N} & \textbf{Clusters} \\
\midrule
HAI-Pre & rule stated & $+$0.372 & 0.276 & 0.178 & 974 & 96 \\
HAI-Pre & conditional cooperation & $-$0.008 & 0.277 & 0.978 & 974 & 96 \\
HAI-Pre & commitment promise & $+$0.464 & 0.277 & 0.094 & 974 & 96 \\
HAI-Pre & cooperation proposal & $+$0.077 & 0.252 & 0.761 & 974 & 96 \\
HAI-Pre & punishment threat & $+$0.097 & 0.251 & 0.701 & 974 & 96 \\
\addlinespace
HAI-Rep & rule stated & $+$0.326 & 0.302 & 0.280 & 1{,}698 & 54 \\
HAI-Rep & conditional cooperation & $-$0.020 & 0.328 & 0.951 & 1{,}698 & 54 \\
HAI-Rep & commitment promise & $+$0.594 & 0.267 & 0.026 & 1{,}698 & 54 \\
HAI-Rep & cooperation proposal & $+$0.209 & 0.352 & 0.554 & 1{,}698 & 54 \\
HAI-Rep & punishment threat & $+$0.430 & 0.295 & 0.145 & 1{,}698 & 54 \\
\addlinespace
HH-Rep & commitment promise & $+$0.430 & 0.390 & 0.270 & 932 & 66 \\
HH-Rep & cooperation proposal & $+$0.523 & 0.382 & 0.171 & 932 & 66 \\
\bottomrule
\end{tabular}
\begin{minipage}{\linewidth}\vspace{2pt}\scriptsize\emph{Notes.} Standard errors clustered at the participant level. Each row is a separate bivariate logit of an indicator for the participant's cooperation in the decision immediately following the message. Predictors are omitted where one cell contains fewer than 20 observations, which excludes conditional cooperation and punishment threats in HH-Rep. The indicator for a stated rule is also omitted in HH-Rep: all 28 such statements are followed by cooperation, so the coefficient is not identified.\end{minipage}
\end{table}

\end{appendices}

\newpage
\bibliographystyle{apacite}
\bibliography{references}
\end{document}